\documentclass[
 reprint,
 amsmath,amssymb,
 aip,
]{revtex4-2}

\usepackage{graphicx}
\usepackage{dcolumn}
\usepackage{bm}
\usepackage{hyperref}
\usepackage{color}
\usepackage{comment}

\begin{document}
\preprint{APS/123-QED}

\title{Density functional study of twisted graphene \\ $L1_0$-FePd heterogeneous interface}

\author{Mitsuharu Uemoto}
\email{uemoto@eedept.kobe-u.ac.jp}
\affiliation{Department of Electrical and Electronic Engineering, Graduate School of Engineering, Kobe University, 1-1 Rokkodai-cho, Nada-ku, Kobe 651-8501, Japan}

\author{Hayato Adachi}
\affiliation{Department of Electrical and Electronic Engineering, Graduate School of Engineering, Kobe University, 1-1 Rokkodai-cho, Nada-ku, Kobe 651-8501, Japan}

\author{Hiroshi Naganuma}
\affiliation{Center for Innovative Integrated Electronics Systems (CIES), Tohoku University, 468-1 Aramaki Aza Aoba, Aoba, Sendai, Miyagi, 980-8572, Japan}
\affiliation{Center for Spintronics Integrated Systems (CSIS), Tohoku University, 2-2-1 Katahira Aoba, Sendai, Miyagi 980-8577 Japan}
\affiliation{Center for Spintronics Research Network (CSRN), Tohoku University, 2-1-1 Katahira, Aoba, Sendai, Miyagi 980-8577 Japan}
\affiliation{Graduate School of Engineering, Tohoku University, \color{blue} 6-6-05, Aoba, Aoba-ku, Sendai, Miyagi, 980-8579, Japan}

\author{Tomoya Ono}
\affiliation{Department of Electrical and Electronic Engineering, Graduate School of Engineering, Kobe University, 1-1 Rokkodai-cho, Nada-ku, Kobe 651-8501, Japan}

\date{\today}

\begin{abstract}
Graphene on $L1_0$-FePd(001), which has been experimentally studied in recent years, is a heterogeneous interface with a significant lattice symmetry mismatch between the honeycomb structure of graphene and tetragonal alloy surface.
In this work, we report on the density functional study of its atomic-scale configurations, electronic and magnetic properties, and adsorption mechanism, which have not been well understood in previous experimental studies.
We propose various atomic-scale models, including simple nontwisted and low-strain twisted interfaces, and analyze their 
energetical stability by performing structural optimizations using the van der Waals interactions of both DFT-D2 and optB86b-vdW functionals.
The binding energy of the most stable structure reached $E_\mathrm{B}=-0.22$~eV/atom for DFT-D2 ($E_\mathrm{B}=-0.19$~eV/atom for optB86b-vdW).
The calculated FePd-graphene spacing distance was approximately 2~\AA, which successfully reproduced the experimental value.
We also find out characteristic behaviors: the modulation of $\pi$-bands, the suppression of the site-dependence of adsorption energy, and the rise of \color{blue} moir\'e-like \color{black} corrugated buckling.
In addition, our atomic structure is expected to help build low-cost computational models for investigating the physical properties of $L1_0$ alloys/two-dimensional interfaces.
\end{abstract}

\maketitle

\section{\label{sec:intro} Introduction}
Iron-palladium \cite{naganuma2015electrical,zhang2018enhancement,klemmer1995magnetic,shima2004lattice,miyata1990ferromagnetic,naganuma2020perpendicular,mohri2001theoretical,iihama2014low,kawai2014gilbert,zharkov2014study,itabashi2013preparation,naganuma2022unveiling} is a binary ordered alloy with a tetragonal $L1_0$ structure that has attracted much attention as a material for spintronic applications \cite{naganuma2015electrical,zhang2018enhancement} because of its high perpendicular magnetic anisotropy (PMA) of $K_u \sim 
10^7~\mathrm{erg}/\mathrm{cm}^3$\cite{klemmer1995magnetic,shima2004lattice,miyata1990ferromagnetic}, and low Gilbert damping of thin films ($\alpha \sim 10^{-2}$) \cite{iihama2014low, kawai2014gilbert}.
These properties are desirable for high-density magnetic random access memory (MRAM)\cite{kent2015new} and magnetic tunneling junction (MTJ)\cite{zhang2018enhancement} devices.
MgO is usually utilized as a barrier layer for FePd-based junctions \cite{parkin2004giant,naganuma2015electrical}; however, the lattice mismatch approaches $10~\%$ which poses a serious problem for a smooth interface and obtaining a high tunnel magnetoresistance (TMR) ratio.
Graphene is considered a potential replacement for MgO layers.
A recent experimental study successfully realized the integration of graphene layers deposited on the $(001)$surface of FePd (FePd$(001)$/Gr) using chemical vapor deposition (CVD) \cite{naganuma2020perpendicular, naganuma2022unveiling}.
It has been reported that graphene protects Fe atoms on the surface from oxidation and provides stable crystallinity, atomic thickness, and flatness control without degrading the perpendicular magnetic properties. 
Graphene grown via CVD is expected to have an energetically stable orientation when formed on tetragonal FePd epitaxial films.
In contrast to the MgO barrier and FePd interface, van der Waals force bonding can be expected to form a flat interface between the graphene and FePd layers.
However, the actual atomic configuration of the C atoms on the surface is not yet well understood.

Graphene on metal surfaces has been intensively studied both experimentally and theoretically, particularly for lattice-matched fcc-metals (\textit{e.g., } Ni(111)/Gr, Cu(111)/Gr, Al(111)/Gr, etc.) \cite{bertoni2005first, kozlov2012bonding, hamada2010comparative, varykhalov2000electronic, abtew2013graphene,mittendorfer2011graphene, Batzill2012}.
Their $(111)$-surface has a threefold symmetry, which is in good agreement with the honeycomb structure of graphene.
C atoms are placed on characteristic "adsorption sites" relative to metallic atoms \cite{bertoni2005first}.
The bonding mechanism of metal-graphene interfaces can be divided into two regimes: physisorption and chemisorption \cite{kozlov2012bonding}.
In the former case, graphene is weakly attracted to metallic atoms via van der Waals (vdW) interactions.
In the latter case, a strong hybridization between the graphene $p\pi$-state and metal $d$-state modifies the electronic band structure and shortens the graphene-metal spacing distance. \textit{ For example }, the typical physisorbed and chemisorbed spacings are estimated as $\sim 3~\mathrm{\AA}$ and $\sim 2$~\AA, respectively \cite{kozlov2012bonding}.
In addition, interfaces with a small lattice constant mismatch (\textit{e.g., } Ir(111)/Gr, Pt(111)/Gr, etc.) exhibit vertical buckling of the graphene, reflecting their \color{blue} moir\'e \color{black} superstructures\cite{Batzill2012,wintterlin2009graphene,hasegawa2013electronic},
where the spacing varies continuously between $2 \sim 3$~\AA.

FePd(001)/Gr exhibited a significant lattice symmetry mismatch. For the tetragonal $L1_0$ structure, Fe (or Pd) atoms on the $(001)$ face are arranged in a square lattice, which is distinct from that of graphene.
The atomic configuration at the interface and the bonding mechanism are not as obvious as those for the fcc-metal surfaces; therefore, theoretical and computational studies are desired.

This study investigates atomic-scale configurations, electronic and magnetic properties, and adsorption mechanisms using density functional theory (DFT)-based calculations. We propose a few designs of the interface structure: simple non-twisted models (simple models) and low-strain twisted models (twisted models). The former is superior in terms of computational costs, and the latter is physically suitable.
We performed structural optimization of various interface models and analyzed their binding and strain energies.
Their influence on the geometrical parameters of graphene (bond length and angle) reveals the reduction of graphene strain energy to obtain a stable structure, and using exhaustive exploration, we find a few possible twisted models. The binding energy of the most stable structure reaches -0.19~eV/atom (-18~kJ/mol), which is of the same order as that of Ni(111)/Gr.
The structural and magnetic properties of our model agree with recent experiments \cite{naganuma2020perpendicular, naganuma2022unveiling}.
In addition, we also find out characteristic behaviors: the modulation of $\pi$-bands, the suppression of the site-dependence of adsorption energy, and the rise of \color{blue} \color{blue} moir\'e \color{black}-like \color{black} corrugated buckling.

The remainder of this paper is organized as follows: in Sec. ~\ref{sec:method}, we present the interface structure models and the computational conditions.
In Sec.~\ref{sec:result}, we report the optimized energies for each structure and discuss their electronic structures.
Finally, in Sec.~\ref{sec:summary}, we present the conclusion.

\section{\label{sec:method} Method}

\begin{figure}
    \includegraphics[width=0.47\textwidth]{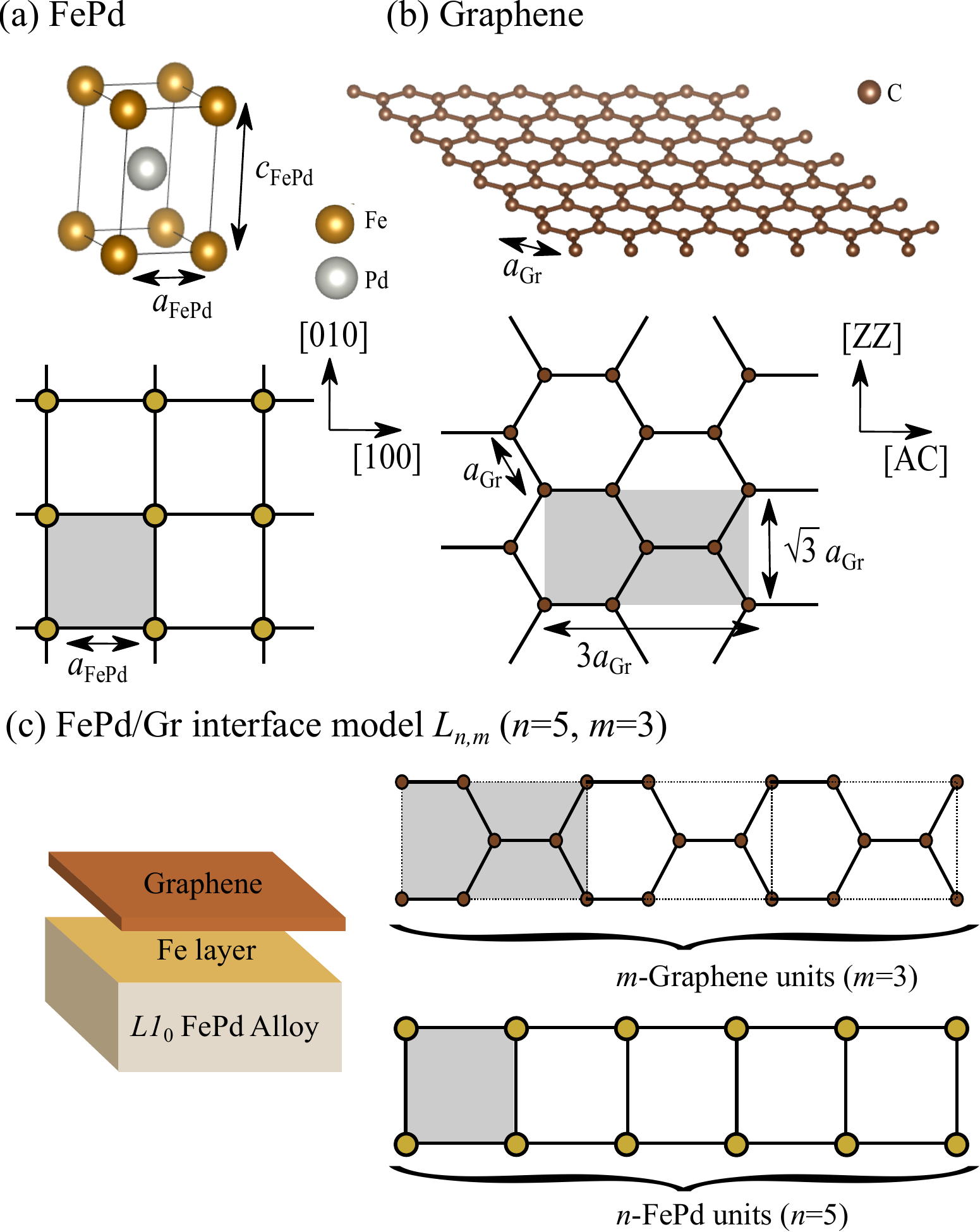}
    \caption{
    Conceptual illustration of simple non-twisted model.
    (a) Bulk $L1_0$-FePd crystal structure and square Fe lattice on the $(001)$ plane.
    (b) Bulk-graphene structure.
    The periodic units of both structures are marked as gray-colored areas.
    (c) $S_{nm}$ interface structure consisting of $n$-FePd units and $m$-graphene periodic units ($n=5, m=3$).
    }
    \label{fig:illust}
\end{figure}

In this study, we propose two atomic-scale structural models of the FePd(001)/Gr interface: a simple non-twisted model (simple model) and a low-strain twisted model (twisted model).
The "simple model" is designed to reduce the number of atoms per supercell, which can significantly reduce the computational time.
However, there is a possibility of a large tensile strain due to the deformation of graphene.
The "twisted model" is a more generalized interface that accounts for twist angles. 
Although it enables a decrease in strain by using adequate supercells, a large computational effort is usually required. 

\subsection{Interface models}
\subsubsection{Simple non-twisted interface model (Simple model)}

Figure~\ref{fig:illust}(a) shows the bulk FePd unit cell with $L1_0$-ordered structure, which contains single Fe and Pd atoms.
The lattice parameters are $a_\mathrm{FePd}=2.67~\mathrm{\AA}$ and $c_\mathrm{FePd}=3.70~\mathrm{\AA}$, which were determined from the structural optimization under equilibrium conditions, in good agreement with previous experimental and theoretical studies \cite{mohri2001theoretical, zharkov2014study, itabashi2013preparation}.
Figure~\ref{fig:illust}(b) illustrates the honeycomb lattice structure of graphene; the gray rectangular area is a single unit cell containing four C atoms in each cell.
The in-plane interatomic distance is $a_\mathrm{Gr} = 1.42~\mathrm{\AA}$.
Owing to the difference in lattice symmetry between FePd and graphene, the atomic interface structures were not obvious.

\begin{figure}
    \includegraphics[width=0.4\textwidth]{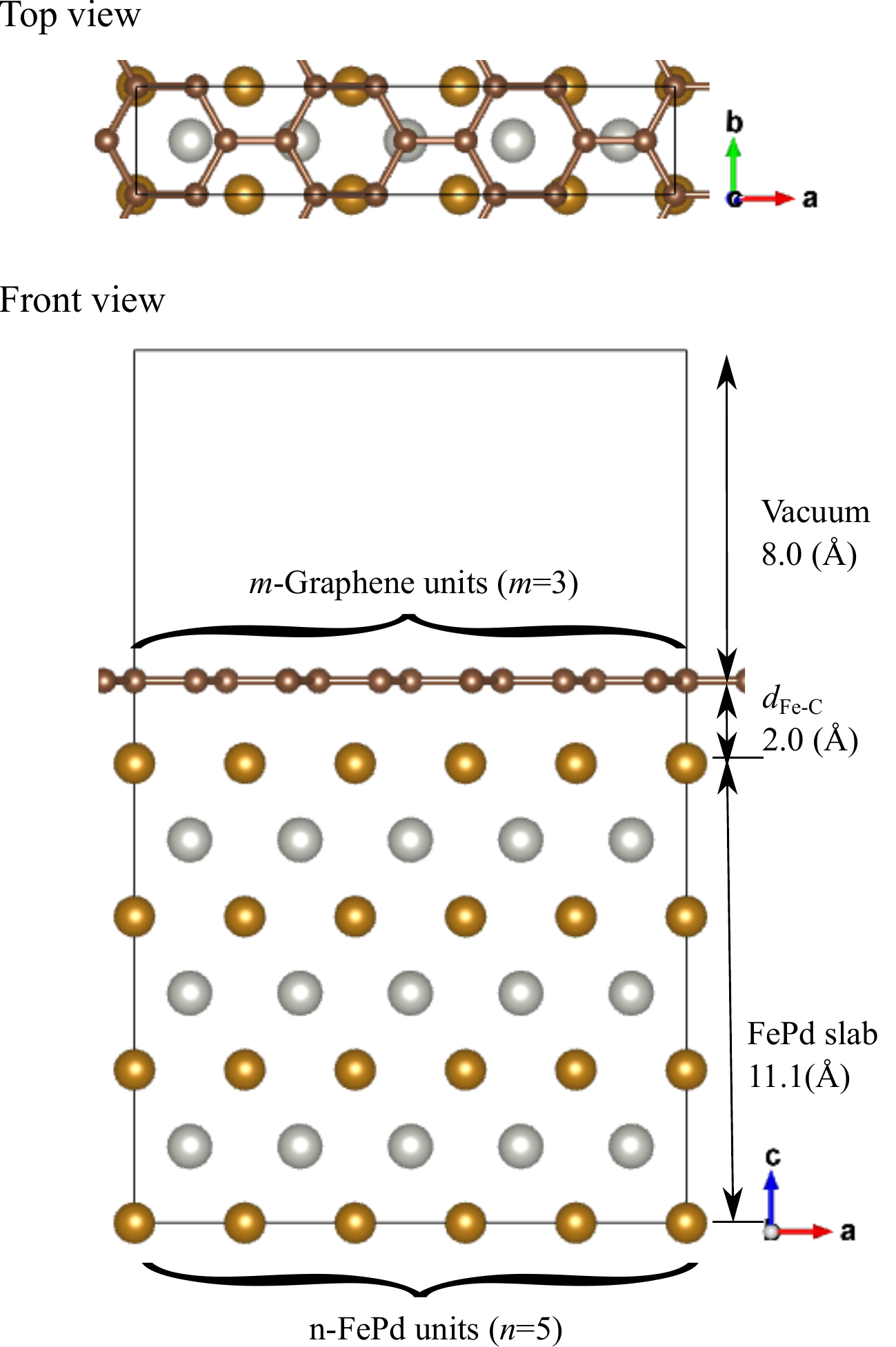}
    \caption{
        Initial structure of $S_{nm}$-type FePd(001)/Gr interface model.
        There are seven atomic layers of FePd, a single graphene layer, and an $8.0~\mathrm{\AA}$ vacuum region. The Fe-C interlayer distance is assumed to be $2~\mathrm{\AA}$ in the initial condition.
    }
    \label{fig:slab}
\end{figure}

As seen Fig.~\ref{fig:illust}(c), the armchair (AC) and zigzag (ZZ) directions of the graphene layer are parallel to the $[100]$ and $[010]$ directions of FePd, respectively.
This structure is constructed based on the following guiding principles:
\begin{enumerate}
\color{blue}
\item  Owing to the attractive interaction between Fe and C atoms, the arrangement in which these atoms are considered stable.
\item The honeycomb structure of graphene on the surface should not have significant distortions.
\end{enumerate}
The period of the graphene ZZ direction is $\sqrt{3}a_\mathrm{Gr} = 2.46~\mathrm{\AA}$. 
However, because the period in the AC direction, $3a_\mathrm{Gr}=4.26~\mathrm{\AA}$, has a significant mismatch with $a_\mathrm{FePd}$, a long-period supercell structure is required in this direction.
We refer to the interface structure consisting of $n$-period FePd units and $m$-period graphene units as $S_{n,m}$ [see Fig. ~\ref{fig:illust}(c)].
This design can be characterized by a pair of integers $n$ and $m$.

To evaluate the stability of the $S_{n,m}$ interface structure, we considered slab supercells as shown in Fig. ~\ref{fig:slab}, which consists of seven atomic layers of FePd with a graphene layer terminating on the top side of the slab.
The bottom two layers were fixed to simulate the bulk properties during structural optimization.
We also used a vacuum layer of approximately $8.0~\mathrm{\AA}$, which was sufficiently large to truncate the interactions under periodic boundary conditions.
The initial spacing between the top Fe layer and graphene ($d_\mathrm{FeC}$) is determined to be $2.0~\mathrm{\AA}$.
The diameter of the $S_{n,m}$ supercell is $ 2.67n~\mathrm{\AA} \times 2.67~\mathrm{\AA} \times 21.1~\mathrm{\AA}$, and the numbers of Fe, Pd, and C atoms are $4n$, $3n$, and $4m$, respectively.

\subsubsection{Low-strain twisted interface model (Twisted model)}

\begin{figure}
    \includegraphics[width=0.45\textwidth]{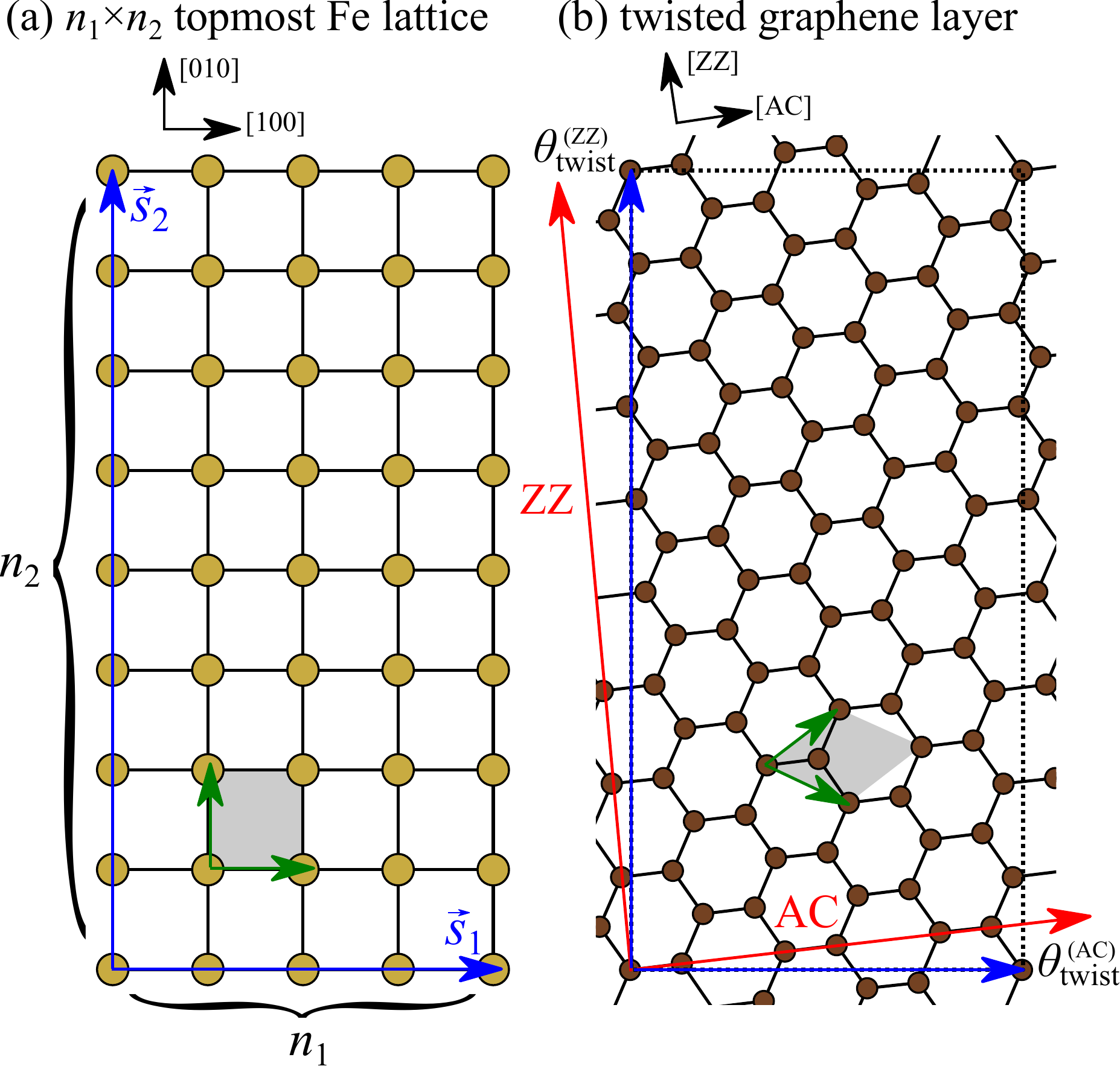}
    \caption{
    Illustration of low-strain twisted interface model represented as $T_{n_1, n_2, m_{11}, m_{12}, m_{21}, m_{22}}$; as an example, $T_{4,8,-5,2,-1,7}$ structure is depicted.
    (a) $n_1 \times n_2$ square lattice of top-most Fe layer.
    (b) twisted graphene layer.
    The green arrows and gray area represent the primitive unit vectors and cell, respectively.
    The blue arrows ($\vec{s}_1$ and $\vec{s}_2$) are supercell lattice vectors, and the red arrows are the AC and ZZ axes.
    }
    \label{fig:twist}
\end{figure}
In the twisted model shown in Fig. ~\ref{fig:twist}, we considered the interface structure \color{blue} with \color{black} the twist angle between FePd and graphene.
The initial arrangement of C atoms was generated by linear transformations involving rotation and resizing onto the original graphene lattice.

Here, we consider the C atom coordinates of freestanding graphene $\vec{r}^\mathrm{(c)}$ as follows:
\begin{align}
\vec{r}^\mathrm{(c)}_{k}
=&
k_1
\vec{a}^\mathrm{(Gr)}_1
+
k_2
\vec{a}^\mathrm{(Gr)}_2
+
\vec{\tau}^\mathrm{(Gr)}_{k_3}
\label{eq:graphene_r}
\end{align}
where $k \equiv (k_1, k_2, k_3)$ are integers,
$\vec{a}^\mathrm{(Gr)}_1$ and $\vec{a}^\mathrm{(Gr)}_2$ are primitive unit vectors of $(3a_\mathrm{Gr}/2,\pm\sqrt{3}a_\mathrm{Gr}/2)$ (green arrows in Fig. ~\ref{fig:twist}(b)), and $\vec{\tau}^\mathrm{(Gr)}_1$ and $\vec{\tau}^\mathrm{(Gr)}_2$ are the sublattice vectors of $(0,0)$ and $(\vec{a}^\mathrm{(Gr)}_1+\vec{a}^\mathrm{(Gr)}_2)/3$, respectively.
In the presence of a twist angle, the transformed C atom coordinates $\vec{r}^{\mathrm{(C)}\prime}_k$ are given by
\begin{align}
\vec{r}^{\mathrm{(C)}\prime}_{k}
=&
\hat{\mathcal{A}}
\vec{r}^\mathrm{(C)}_{k}
\end{align}
with $2 \times 2$ transform matrix, $\hat{\mathcal{A}}$ (Affine map).
As seen in Fig.~\ref{fig:twist}(a), we assume a supercell including $n_1 \times n_2$ square Fe lattice; the unit vectors are represented as $\vec{s}_1=n_1 \vec{a}^\mathrm{FePd}_1$ and $\vec{s}_2=n_2 \vec{a}^\mathrm{FePd}_2$ (blue arrows) with $\vec{a}^\mathrm{FePd}_1=(a_\mathrm{FePd},0)$ and $\vec{a}^\mathrm{FePd}_2=(0, a_\mathrm{FePd})$ (green arrows in Fig. ~\ref{fig:twist}).
Because the Fe and C layers have the same periodicity, the following conditions are required:
\begin{align}
\left\{
\begin{aligned}
\vec{s}_1
=&
m_{11} \hat{\mathcal{A}}\vec{a}^\mathrm{(Gr)}_1
+
m_{12} \hat{\mathcal{A}} \vec{a}^\mathrm{(Gr)}_2
\\
\vec{s}_2 
=
& 
m_{21} \hat{\mathcal{A}} \vec{a}^\mathrm{(Gr)}_1
+
m_{22} \hat{\mathcal{A}} \vec{a}^\mathrm{(Gr)}_2
 \end{aligned}
 \right.
 \;,
\end{align}
where $m_{11}$, $m_{12}$, $m_{21}$., and $m_{22}$ are the integers.
By solving the linear system of equations, we obtain 
\begin{align}
\left\{
\begin{aligned}
 \hat{\mathcal{A}} \vec{a}^\mathrm{(Gr)}_1
 =&
 \mu_{11} n_1 \vec{a}^\mathrm{Fe}_1
 +
 \mu_{12} n_2 \vec{a}^\mathrm{Fe}_2
 \\
 \hat{\mathcal{A}} \vec{a}^\mathrm{(Gr)}_2
 =&
 \mu_{21} n_1 \vec{a}^\mathrm{Fe}_1
 +
 \mu_{22} n_2 \vec{a}^\mathrm{Fe}_2
 \end{aligned}
 \right.
\end{align}
with
\begin{align}
\begin{pmatrix}
    \mu_{11} & \mu_{12} \\
    \mu_{21} & \mu_{22}
\end{pmatrix}
=&
\begin{pmatrix}
    m_{11} & m_{12} \\
    m_{21} & m_{22}
\end{pmatrix}^{-1}
\;.
\end{align}
Therefore, we have
\begin{align}
    \hat{\mathcal{A}}
    =&
    \left(
 \mu_{11} n_1 \vec{a}^\mathrm{Fe}_1
 +
 \mu_{12} n_2 \vec{a}^\mathrm{Fe}_2
    \right) \left(\vec{b}^\mathrm{(Gr)}_1 \right)^\mathrm{T}
    \notag \\ & +
    \left(
 \mu_{21} n_1 \vec{a}^\mathrm{Fe}_1
 +
 \mu_{22} n_2 \vec{a}^\mathrm{Fe}_2
    \right)
    \left(
        \vec{b}^\mathrm{(Gr)}_2
    \right)^\mathrm{T}
\end{align}
where $\vec{b}^\mathrm{(Gr)}$ are reciprocal vectors of $\vec{a}^\mathrm{(Gr)}$.
The above formalism provides a universal description of the twisted interface, which can be characterized by six integers: ${n_1, n_2, m_{11}, m_{12}, m_{21}, m_{22}}$.
Therefore, we represent the structure as $T_{n_1, n_2, m_{11}, m_{12}, m_{21}, m_{22}}$;
as an example, Fig.~\ref{fig:twist}(a)–(b) shows $T_{4,8,-5,2,-1,7}$ model.
Similar to Fig.~\ref{fig:slab}, slab supercell structures containing the twisted graphene layer on seven atomic layers of FePd were employed in the calculations.

\subsection{Computation}

For computation, we performed density functional theory (DFT) calculations using the Vienna ab initio simulation package (VASP) 6.2, which uses the projector augmented wave (PAW) method \cite{blochl1994projector} and a plane-wave basis set.
We also employed the generalized gradient approximation (GGA) exchange-correlation functional of the Perdew–Burke–Ernzerhof (PBE) \cite{perdew1996generalized} with Grimme's DFT-D2 method for vdW interactions \cite{grimme2006semiempirical}.
Besides, for comparison, we also use optB86b-vdW functional \cite{klimes2009, klimes2011}.
We used a plane-wave basis cutoff energy of $400~\mathrm{eV}$. 
The first Brillouin zone was sampled using Gamma-centered $k$-point grids.
\color{blue}
The magnetization is treated by spin-collinear calculations.
\color{black}
In the case of the simple model $S_{n,m}$, we used $n_{k} \times 9 \times 1$ grids, where $n_{k}$ is the smallest integer satisfying $n_{k} \times n \geq 9$.
For the twisted model, we used $n_{k_1} \times n_{k_2} \times 1$ with $n_{k_i} \times n_i \geq 9$.

\section{\label{sec:result} Result and Discussion}

Here, we consider a few different "simple model" interface structures ($S_{3,2}$, $S_{4,2}$, $S_{4,3}$, $S_{5,3}$, $S_{5,4}$, and $S_{7,4}$).
To determine their stability, we introduce the binding energy $E_\mathrm{B}$ using the following formula:
\begin{align}
    E_\mathrm{B}(n,m)
    =&
    \frac{
        E_\mathrm{FePd/Gr}(n, m)
        -
        E_\mathrm{FePd}(n)
        -
        E_\mathrm{Gr}(m)
    }{n_\mathrm{C}(m)}
    \label{eq:binding}
\end{align}
with the number of C atoms $n_\mathrm{C}(m) = 4m$, where $E_\mathrm{FePd/Gr}(n, m)$ is the total energy of the relaxed FePd(001)/Gr slab structure and $E_\mathrm{FePd}(n)$ and $E_\mathrm{Gr}(m)$ are that of the pristine FePd slab and the pristine free-standing graphene in the equilibrium conditions, respectively.
Additionally, we calculated the strain energy $E_\mathrm{S}$ as follows:
\begin{align}
E_\mathrm{S}(n, m)
=&
\frac{\tilde{E}_\mathrm{Gr}(n,m)-E_\mathrm{Gr}(m)}{n_\mathrm{C}(m)}
\;,
\end{align}
where $\tilde{E}_\mathrm{Gr}(n,m)$ is the energy of deformed graphene with the same C atom coordinates as $E_\mathrm{FePd/Gr}(n,m)$.

\begin{table*}
\caption{
\label{tbl:simple}
Geometrical parameters of the initial atomic configuration of the simple models ($S_{n,m}$) and the calculated energies of the relaxed structures:
the mean absolute error (MAE) of the C atoms' bond length $a_\mathrm{C}$ and angle $\theta_\mathrm{C}$ from the equilibrium free-standing graphene ($1.42~\text{\AA}$ and $120^\circ$) are indicated as MAE~$a_\mathrm{C}/a_\mathrm{Gr}$ and MAE~$\theta_\mathrm{C}$.
The binding energies $E_\mathrm{B}$, strain energies $E_\mathrm{S}$, and spacing distances $d_\mathrm{FeC}$ are calculated by both of DFT-D2 and optB86b-vdW (the values in parentheses) functionals.
}
\begin{ruledtabular}
\begin{tabular}{llcccrrr}
&
Model & 
$n/m$ &
$\mathrm{MAE}~a_\mathrm{C}/a_\mathrm{Gr}$ &
$\mathrm{MAE}~\theta_\mathrm{C}$ &
$E_\mathrm{B}$~(eV/atom) &
$E_\mathrm{S}$~(eV/atom) &
$d_\mathrm{FeC}$~(\AA)\\ 
\hline
a & $S_{3,2}(\sim S_{6,4})$ & $1.50$ & $15~\%$ & $6^\circ$ & $-0.02~(+0.01)$ & $+0.25~(+0.25)$ & 2.04~(2.04) \\ 
b & $S_{4,2}(\sim S_{2,1}, S_{6,3})$ & $2.00$ & $28~\%$ & $6^\circ$ & $+0.24~(+0.25)$ & $+1.47~(+1.47)$ & N/A \\ 
c & $S_{4,3}$ & $1.33$ & $16~\%$ & $10^\circ$ & $+0.27~(+0.29)$ & $+0.68~(+0.67)$ & N/A \\
d & $S_{5,3}$ & $1.67$ & $16~\%$ & $2^\circ$ & $-0.11~(-0.09)$ & $+0.25~(+0.25)$ & $1.98~(1.99)$ \\
e & $S_{5,4}$ & $1.25$ & $17~\%$ & $12^\circ$ & $+0.38~(\color{blue} +0.27)$ & $+0.63~(+0.63)$ & N/A \\
f & $S_{7,4}$ & $1.75$ & $19~\%$ & $0.4^\circ$ & $+0.00~(+0.03)$ & $+0.44~(+0.44)$ & 1.96~(1.96)
\end{tabular}
\end{ruledtabular}
\end{table*}

Table~\ref{tbl:simple} shows $E_\mathrm{B}$ and $E_\mathrm{S}$ values of $S_{n,m}$ models, and for comparison, we tested both the DFT-D2 and optB86b-vdW functionals.
In the DFT-D2 results, most of the structures have a positive $E_\mathrm{B}$ value, which is energetically unstable, and only two structures have a negative $E_\mathrm{B}$.
$S_{5,3}$ is the most stable, with $E_\mathrm{B}=-0.11$~eV/atom.
In the optB86-vdW results, $S_{5,3}$ is also stable, with $E_\mathrm{B}=-0.09$~eV/atom.
$S_{5,3}$ is the structure with the smallest $E_\mathrm{S}=0.25$~eV/atom, and the magnitude of $E_\mathrm{S}$ is comparably large to $E_\mathrm{B}$.
We expect that strain in graphene is essential for the stability of these simple models.
\begin{figure}
    \includegraphics[width=0.45\textwidth]{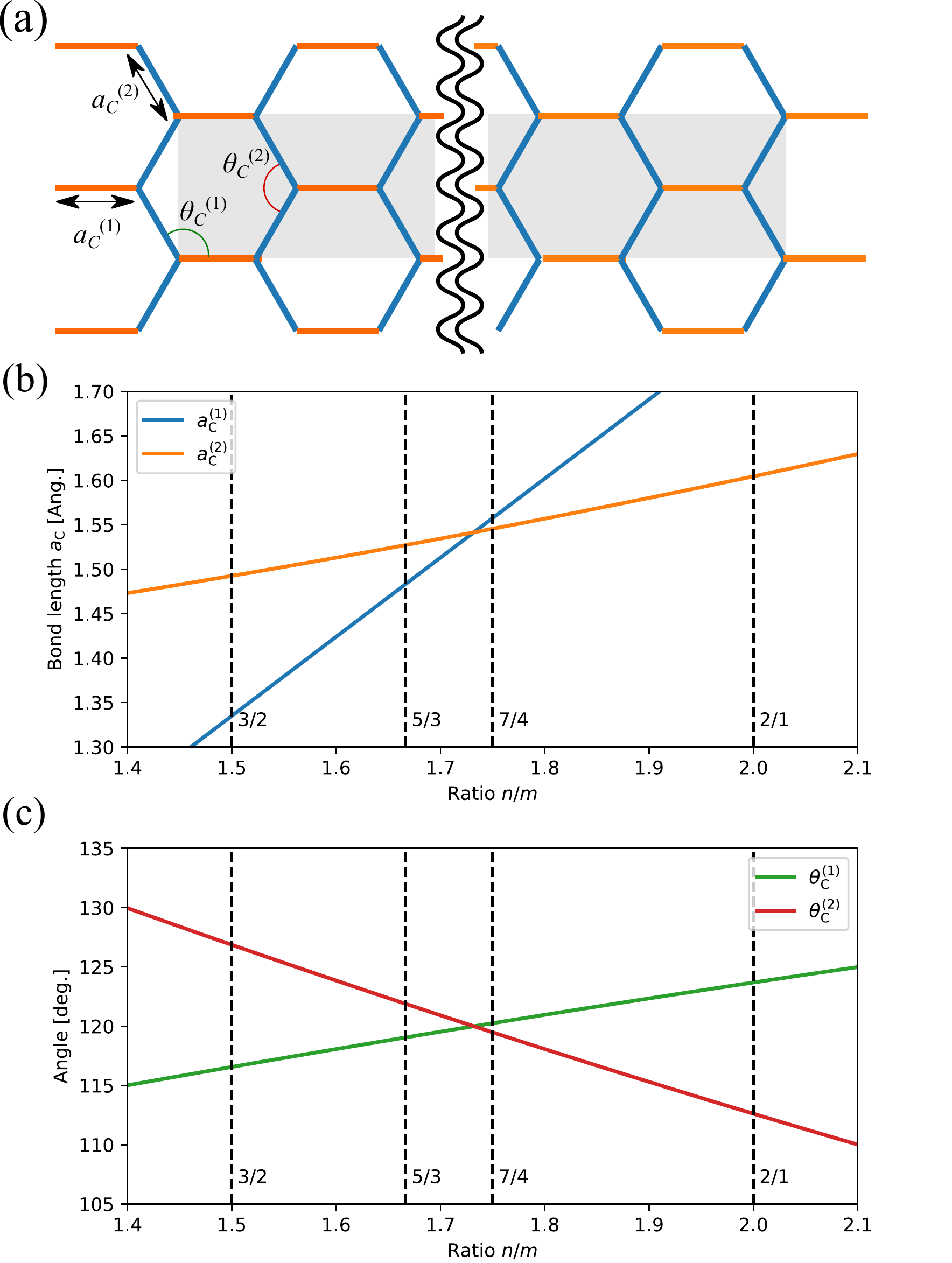}
    \caption{
        Estimated modulation of the bond length and bond angle of the graphene layer in the $S_{nm}$ structure.
        (a) Definition of bond length $a^{(1)}_\mathrm{C}$ , $a^{(2)}_\mathrm{C}$, and bond angle $\theta^{(1)}_\mathrm{C}$ , $\theta^{(2)}_\mathrm{C}$.
        (b) $a^{(1)}_\mathrm{C}$ and $a^{(2)}_\mathrm{C}$ as functions of the $n/m$ ratio.
        (c) $\theta^{(1)}_\mathrm{C}$ and $\theta^{(2)}_\mathrm{C}$ as functions of the $n/m$ ratio.
    }
    \label{fig:modulation}
\end{figure}
\color{blue}
Moreover, $d_\mathrm{FeC}$ in Table~\ref{tbl:simple}
represents a distance between the topmost Fe and C layer.
For some of the energetically unstable ($E_\mathrm{B} \gg 0$) interface models, C atoms are not adsorbed on metal surfaces, or not possible to construct a flat atomic surface. 
For example, in the case of $S_{5,4}$, the vertical distances from the Fe surface to the nearest / farthest C atom are widely distributed between $1.5 \sim 4.2$~\AA. [See S1 in the supplementary material]
Therefore, for a few unstable structures, $d_\mathrm{FeC}$ values in Table~\ref{tbl:simple} are indicated as N/A.
\color{black}

We then consider the bond length and bond angle distributions in the initial state of the optimization.
Here, we consider the initial coordinates of C atoms in $S_{n,m}$. The bond lengths are represented by two lengths $a_\mathrm{C}^{(1)}$ and $a_\mathrm{C}^{(2)}$, as shown in Fig. ~\ref{fig:modulation}(a);
$a_\mathrm{C}^{(1)}$ and $a_\mathrm{C}^{(2)}$ can be analytically expressed as functions of the ratio $n/m$:
\begin{align}
a_\mathrm{C}^{(1)}
=&
\frac{1}{3} \left(\frac{n}{m}\right) a_\mathrm{FePd}
\\
a_\mathrm{C}^{(2)}
=&
\sqrt{\frac{1}{4} + \frac{1}{36}\left(\frac{n}{m}\right)^2} a_\mathrm{FePd}
\;.
\end{align}
In Fig.~\ref{fig:modulation}(b), we plotted the $n/m$ dependence of $a_\mathrm{C}^{(1)}$ and $a_\mathrm{C}^{(2)}$.
As the equilibrium bond length of graphene is approximately $a_\mathrm{Gr}=1.42~\mathrm{\AA}$, \color{blue} to reduce \color{black} the strain, the value of $n/m$ should be approximately $\approx 1.5$.
Using a similar procedure, the bond angles $\theta_\mathrm{C}^{(1)}$ and $\theta_\mathrm{C}^{(2)}$ shown in Fig. ~\ref{fig:modulation}(a), can be calculated using the following equation:
\begin{align}
\theta_\mathrm{C}^{(1)}
=&
\cos^{-1} \left[
    - \frac{(n/m)}{6 \sqrt{1/4+(1/36)(n/m)^2}}
\right]
\\
\theta_\mathrm{C}^{(2)}
=&
2 \cos^{-1} \left[
    \frac{(n/m)}{6 \sqrt{1/4+(1/36)(n/m)^2}}
\right]
\;.
\end{align}
Figure3(c) shows the n/m dependence of $\theta_\mathrm{C}^{(1)}$ and $\theta_\mathrm{C}^{(2)}$;
the angle approaches $120^\circ$ at approximately $n/m \approx 1.7$.
This result indicates that the distortion of the graphene lattice is minimized in the n/m range of $1.5 \sim 1.7$, which is consistent with Table~\ref{tbl:simple}.
This trend can be attributed to the strain in the honeycomb structure of graphene.

\begin{figure}
    \centering
    \includegraphics[width=0.33\textwidth]{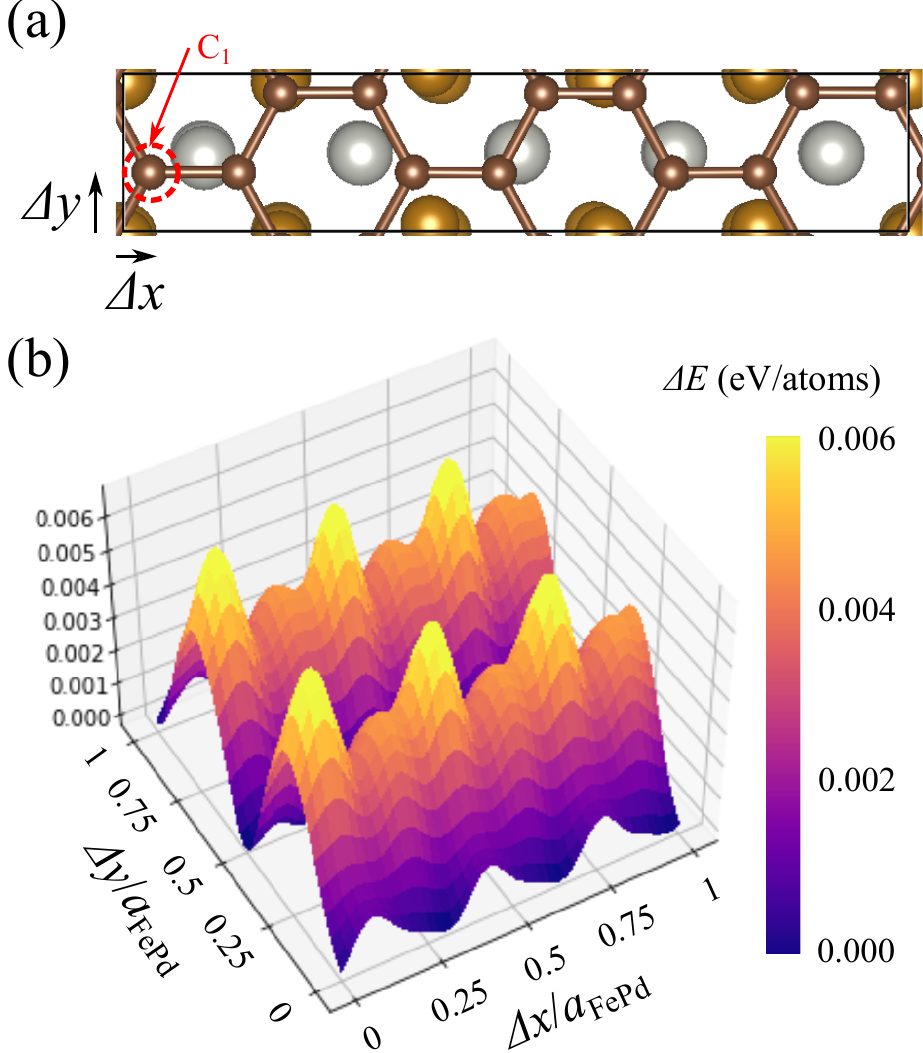}
    \caption{
        \label{fig:shift}
        Potential energy surface for lateral shifts of the graphene layer on FePd(001).
        (a) Schematic illustration. $\Delta{x}$ and $\Delta{y}$ represent the displacements of the $\mathrm{C}_1$ atom from the top of a single Fe site.
        (b) Dependence of the binding energy difference $\Delta{E}$ on $(\Delta{x}, \Delta{y})$.
        \color{blue}
        ($S_{5, 3}$-type model is employed in this calculation)
    }
\end{figure}
Next, we calculated the potential energy surface (PES) profile: adsorption-energy dependency 
on the lateral shift of the graphene layer with respect to that of the topmost Fe layer.
Figure~\ref{fig:shift}(a) shows a schematic illustration.
$\Delta{x}$ and $\Delta{y}$ represent the magnitudes of the displacements of $\mathrm{C}_1$ atom from Fe atom.
Structural optimizations were performed by fixing the lateral position of the $\mathrm{C}_1$ atom and obtaining the binding energy as a function of displacement $E_\mathrm{B}(n, m; \Delta{x}, \Delta{y})$.
\color{blue}
We use $S_{5,3}$-type interface model.
\color{black}
Figure~\ref{fig:shift}(b) plots $\Delta{E}(\Delta{x},\Delta{y})$ defined as:
\begin{align}
    \Delta{E}(\Delta{x},\Delta{y})
    =&
    E_\mathrm{B}(n,m;\Delta{x},\Delta{y})
    -
    E_\mathrm{B}(n,m;0,0)
    \;.
\end{align}
Equivalent points appear periodically in the PES profile because of the translational symmetry of the system.
An energetically stable minimum appears in the vicinity of $(\Delta{x},\Delta{y})=(0, 0)$, which corresponds to the case when a single C atom is at the top of the Fe site, and $\Delta{E}$ becomes unstable when every C atom is equally distant from the Fe atoms.
This sub-$a_\mathrm{Fe}$ scale change is a result of the attractive short-range interaction between Fe and C atoms, which reflects chemisorption.
In contrast, $\Delta{E}$ did not exceed $0.006~\mathrm{eV/atom}$, which is less than 5\% of the binding energy.
This 5~\% change is below the typical value obtained for a chemisorbed interface such as Ni(111)/Gr \cite{kozlov2012bonding}.
This behavior was similar to that of the physisorbed case;
\color{blue}
it will be also discussed in the last section of this paper.
\color{black}

\begin{figure}
    \includegraphics[width=0.4\textwidth]{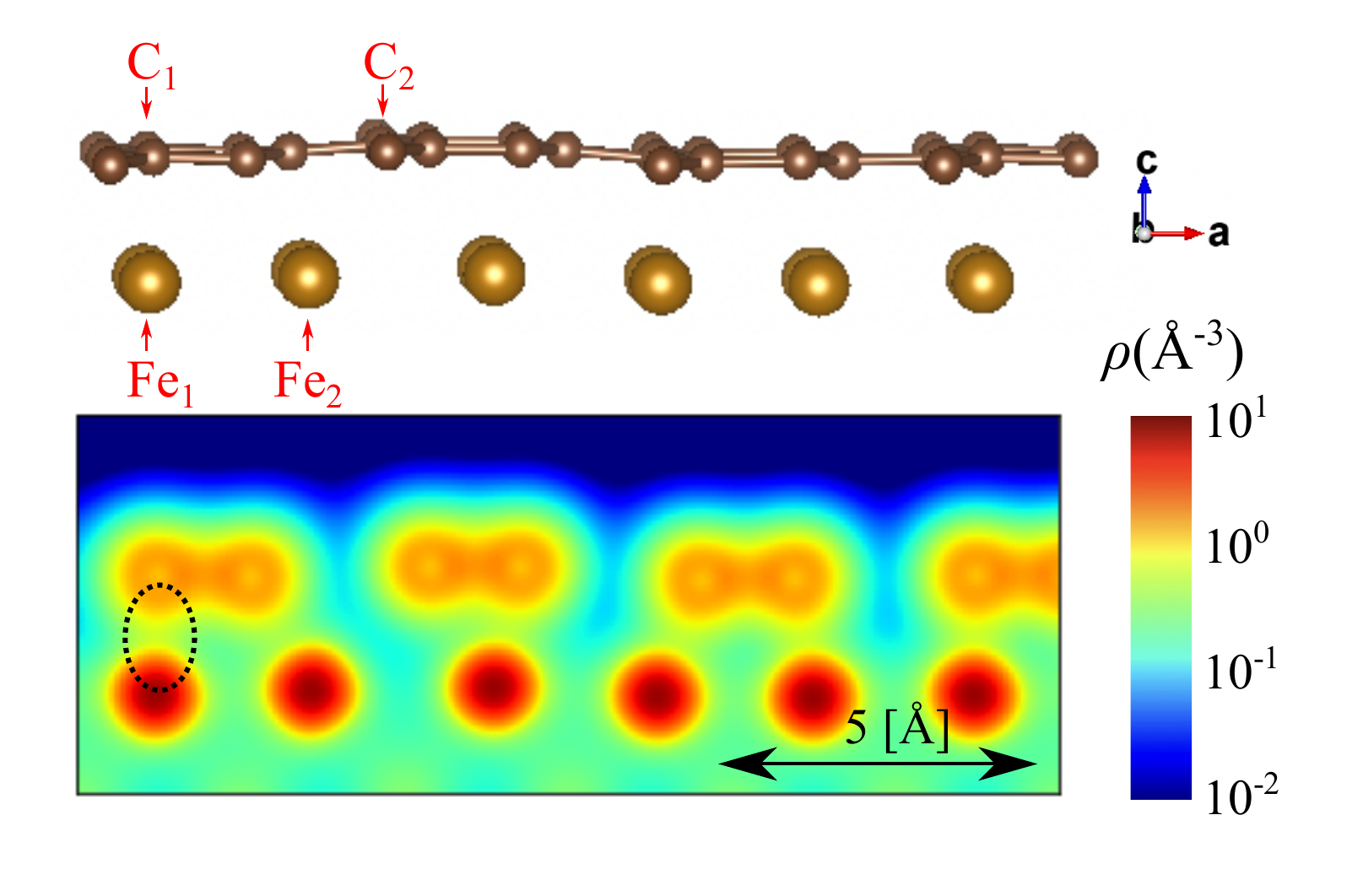}
    \caption{
        Charge density profile of the $S_{5,3}$-type FePd(001)/Gr interface.
        $\mathrm{C}_1$ and $\mathrm{Fe}_1$ represent the closest pair, and $\mathrm{C}_2$ and $\mathrm{Fe}_2$ are examples of sites with no close C or Fe atoms.
    }
    \label{fig:rho}
\end{figure}

Figure~\ref{fig:rho} shows the electron-density profile of the $S_{5,3}$ structure.
A high-density area is found between labeled $\mathrm{C}_1$ and the neighboring C atom, which represents a C-C $\mathrm{sp}^2\sigma$ bond orbital.
Although most $\mathrm{C}-\mathrm{Fe}$ bonds are not seen as $\mathrm{C}-\mathrm{C}$, there is an area of slightly higher density between $\mathrm{C}_1$ and $\mathrm{Fe}_1$ (displayed by the broken circle), which is one of the closest Fe-C pairs.
In addition, the average Fe-C binding distance is approximately $2~\mathrm{\AA}$, which is a typical value for chemisorbed metal/Gr interfaces \cite{kozlov2012bonding, piquemal2020spin}.

\begin{table*}
\caption{\label{tbl:twist}
Geometrical parameters and energies of the twisted FePd(001)/Gr models.
The size of supercell $n_1 \times n_2$, and the twisted angle $\theta^\mathrm{(AC)}_\mathrm{twist}$ and $\theta^\mathrm{(ZZ)}_\mathrm{twist}$ are explained in Fig.~\ref{fig:twist}.
$n_\mathrm{C}$ is the number of C atoms in the supercell.
similar to the table~\ref{tbl:simple},
$E_\mathrm{B}$, $E_\mathrm{S}$, and $d_\mathrm{FeC}$ represents the binding energy, strain energy and the spacing distance calculated by using DFT-D2 and optB86b-vdW (the values in parentheses) functionals.
}
\begin{ruledtabular}
\begin{tabular}{cccccccrrr}
&
Model & 
$n_1 \times n_2$ &
$n_\mathrm{C}$ &
$\theta_\mathrm{twist}^\mathrm{(AC)},
\theta_\mathrm{twist}^\mathrm{(ZZ)}$ &
$\mathrm{MAE}~a_\mathrm{C}/a_\mathrm{Gr}$ &
$\mathrm{MAE}~\theta_\mathrm{C}$ &
$E_\mathrm{B}$~(eV/atom) &
$E_\mathrm{S}$~(eV/atom) & 
$d_\mathrm{FeC}$~(\AA)
\\ 
\hline
g &
$T_{4,6,-5,2,-1,7}$  & 
$4 \times 6$ & $66$ & $6.5^\circ$, $7.5^\circ$ &
$0.5~\%$ & $0.5^\circ$ & 
$-0.20~(-0.16)$ &
$0.02~(0.02)$ &
$2.11~(2.10)$
\\
h &
$T_{4,7,-5,2,-1,8}$  & 
$4 \times 7$ & $76$ & $6.8^\circ$, $6.4^\circ$ &
$0.3~\%$ & $0.3^\circ$ & 
$-0.21~(-0.18)$ & 
$0.01~(0.01)$ &
$2.08~(2.08)$
\\
i &
$T_{4,8,-5,2,-1,8}$  & 
$4 \times 8$ & $86$ & $6.9^\circ$, $5.7^\circ$ &
$0.5~\%$ & $0.8^\circ$ & 
$-0.22~(-0.19)$ & 
$0.02~(0.02)$ &
$2.07~(2.07)$
\end{tabular}
\end{ruledtabular}
\end{table*}

\begin{figure}
    \includegraphics[width=0.4\textwidth]{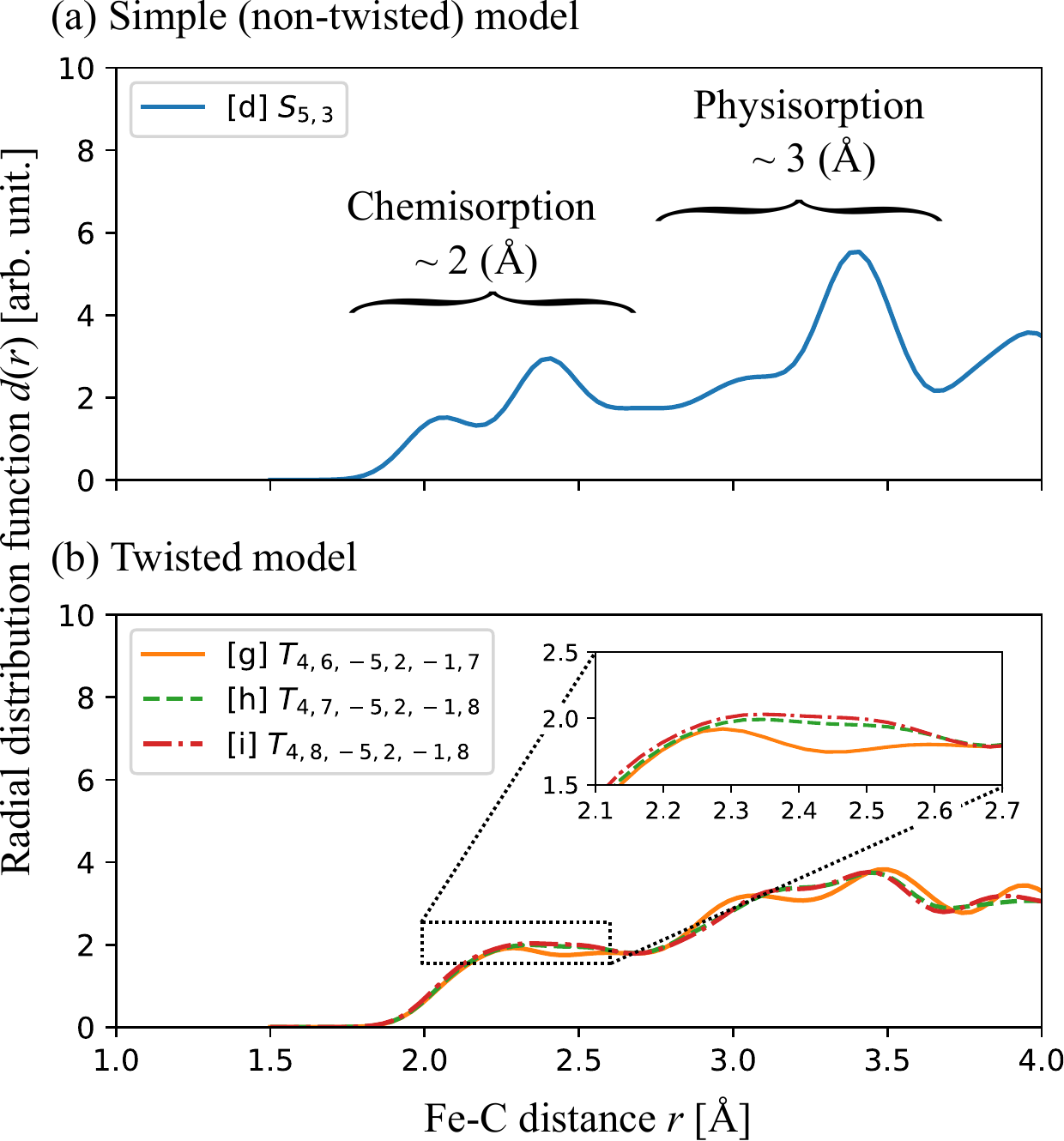}
    \caption{
    Radial distribution function of the C-Fe bonds for the stable structures of simple model (a) and twisted models (b).
    The typical distances for chemisorption ($\sim 2~\mathrm{\AA}$) and physisorption ($\sim 3~\mathrm{\AA}$) are labeled for convenience.
    }
    \label{fig:radial}
\end{figure}

\begin{figure}
    \includegraphics[width=0.4\textwidth]{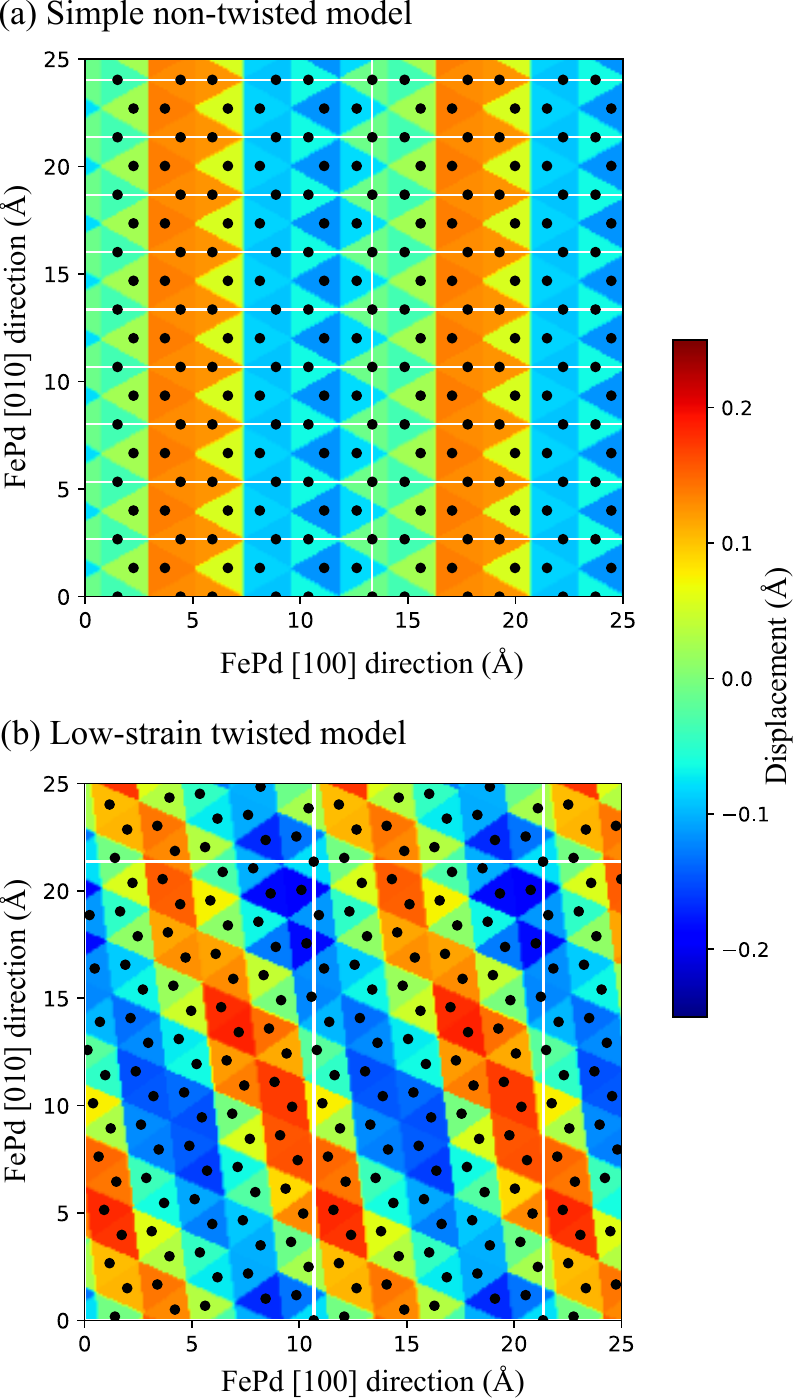}
    \caption{
        \label{fig:height}
        Height profile of the Gr layer of optimized simple model $S_{5,3}$ (a) and the twisted model $T_{4,8,-5,2,-1,8}$ (b).
        The black points represent lateral ($xy$-) position of C atoms, and the colormaps show their vertical ($z$-) position displacement from the average level.
        \color{blue}
        The white lines indicates the periodicity of the supercells.
    }
\end{figure}

\begin{figure*}
    \includegraphics[width=0.95\textwidth]{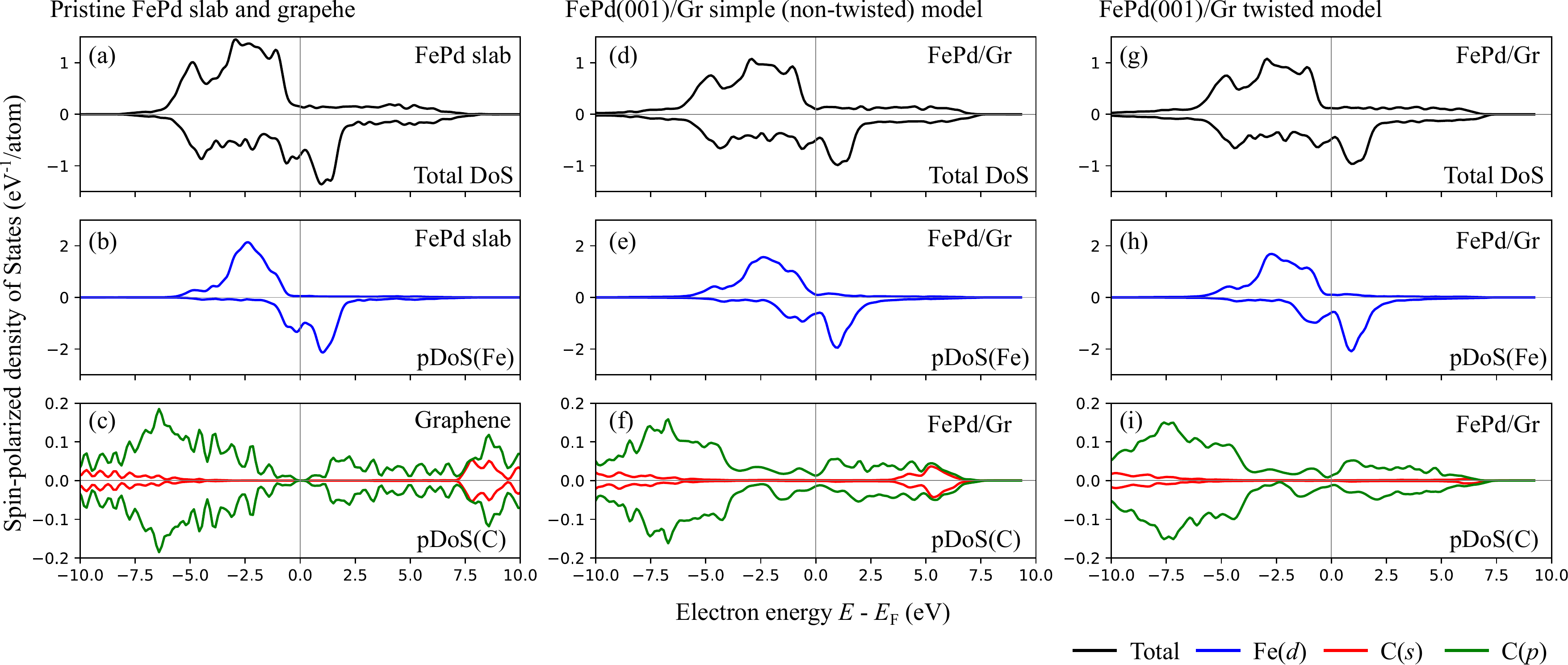}
    \caption{
        \label{fig:dos}
        \color{blue} Spin-polarized \color{black} density of states (DoS) and partial DoS (pDoS) profiles:
        (a) DoS for the pristeine FePd
        (b) pDoS of topmost Fe(3d) for the pristeine FePd
        (c) pDoS of C(s) and C(p) for free-standing graphene.
        (d-f) DoS and pDoS for FePd(001)/Gr interface of the simple $S_{5,3}$ model [illustrated in Fig.~\ref{fig:illust}(c)]. 
        (g-i) FePd(001)/Gr interface of the twisted $T_{4,8,-5,2,-1,8}$ model [illustrated in Fig.~\ref{fig:twist}].
    }
\end{figure*}

Next, we consider low-strain twisted interface models.
The twisted model $T_{n_1, n_2, m_{11}, m_{12}, m_{21}, m_{22}}$ has six individual parameters and numerous structures.
We exhaustively searched for every possible set of $n_1$, $n_2$, $m_{11}$, $m_{12}$, $m_{21}$, and $m_{22}$ with the restriction that supercell size $n_1 \times n_2$ is smaller than $9 \times 9$.
From the analysis of $S_{nm}$ model, we assumed that a screening condition that reduces the strain in the initial structure is essential for stability.
In Table~\ref{tbl:twist}, we selected three structures with minimal MAE~$a_\mathrm{C}/a_\mathrm{Gr}$: $T_{4,6,-5,2,-1,7}$, $T_{4,7,-5,2,-1,8}$ and $T_{4,8,-5,2,-1,8}$ models, for convenience, which are also represented as $\text{g}$, $\text{h}$, and $\text{i}$, respectively.
Model $\text{i}$, illustrated in Fig. ~\ref{fig:twist}, is the most stable structure among the three cases.
$E_B$ was $-0.22$~eV/atom ($\approx -21$~kJ/mol) in DFT-D2 and $-0.19$ ($\approx-18$~kJ/mol) in optB86b-vdW.
Because this value is of the same order as that of the well-known chemisorbed Ni(111)/Gr ($\approx -12$~kJ/mol) \cite{kozlov2012bonding}, the adsorption state is considered very stable.
The above twisted models can dramatically decrease strain; in model $\text{g}$, $\text{h}$ and $\text{i}$, MAE~$a_\mathrm{C}/a_\mathrm{Gr}$ is lesser than 0.5~\%, and $E_\mathrm{S}$ is $0.01 \sim 0.02$~eV/atom.
Although $\mathrm{MAE}~a_\mathrm{C}/a_\mathrm{Gr}$ and $\mathrm{MAE}~\theta_\mathrm{C}$ became larger in the order of model $\text{g} < \text{h} < \text{i}$, model $\text{i}$ is the most stable amongst the three. 
In such a low-strain model, the Fe-C interaction is more essential than MAEs.

Next, we considered the radial distribution function of C-Fe bonds, which is defined as follows:
\begin{align}
d(r) &= \frac{1}{n_\mathrm{c}} \sum_{k}^\mathrm{(Fe)} \sum_{k'}^\mathrm{(C)} \delta\left(r - \| \vec{r}^\mathrm{(Fe)}_{k} - \vec{r}^\mathrm{(C)}_{k'} \| \right)
\;,
\end{align}
where $\vec{r}^\mathrm{(Fe)}_{k}$ and $\vec{r}^\mathrm{(C)}_{k'}$ represent the coordinates of $k$ and $k'$-th topmost Fe and C atoms, respectively. 
For computational convenience, we used a Gaussian of $\sigma=0.1~\mathrm{\AA}$ as the $\delta$-function above.
Figure~\ref{fig:radial}(a) shows $d(r)$ of the stable $S_{5,3}$ model (d in Table. ~\ref{tbl:simple}). $d(r)$ has a broad band from $2~\mathrm{\AA}$ to $3~\mathrm{\AA}$.
Generally, the spacing of the typical chemisorbed fcc-metal/Gr interface is approximately $2~\mathrm{\AA}$ and that of the physisorbed interface is approximately $3~\mathrm{\AA}$.
In the case of FePd(001)/Gr, the averaging of the interatomic interactions between C and Fe resulted in an intermediate behavior between chemisorption and physisorption.
Similar behavior was observed in the twisted model [see Fig.~\ref{fig:radial}(b)]. 
d(r) around $r=2$~\AA \; \color{blue} becomes \color{black} slightly larger in the order of the model $\text{g} < \text{h} < \text{i}$, representing the number of Fe-C pairs at a typical chemisorption distance.
This trend is consistent with $E_\mathrm{B}$ in Table~\ref{tbl:twist}; in the low-strain case, Fe-C inter-atomic interaction is also essential.

As seen in Fig.~\ref{fig:rho}, the height of the C atoms is not flat but rather buckled vertically.
Generally, in the typical fcc-metal/Gr case, buckled graphene is often observed at the interface with a \color{blue} moir\'e \color{black} structure by small lattice constant mismatches, such as Ir(111)/Gr and Pt(111)/Gr. Their buckling amplitudes are usually more than $1~\text{\AA}$ \cite{Batzill2012} and the \color{blue} moir\'e \color{black} pattern because of the long-period (more than a few nanometers) hexagonal lattice \cite{wintterlin2009graphene}.
For FePd(001)/Gr, the colormaps in Fig. ~\ref{fig:height} shows the height profile of the graphene layer in the optimized structures. Surprisingly, well-defined vertical corrugations of approximately $0.4 \sim 0.5~\text{\AA}$ were observed.
Unlike the \color{blue} moir\'e \color{black} pattern, the model in our present study has a stripe-like pattern instead of a hexagonal pattern owing to the lattice symmetry mismatch.
Some C atoms, which are close to the top Fe sites, feel a relatively stronger attractive interaction than any other C atoms.
This inhomogeneity results in buckling, but their amplitudes are largely suppressed by averaging the interatomic interactions.
The C atom at the origin in Fig. ~\ref{fig:height}, which is situated on top of the Fe sites, is in the valley region of the corrugation.
\color{blue}

\color{black}

Figure~\ref{fig:dos} presents the density of states (DoS) and partial DoS (pDoS) profiles.
The plots (a)–(c) represent DoS and pDoS for pristine FePd and graphene. Figure~\ref{fig:dos} (d)–(f) and Figure~\ref{fig:dos} (g)–(i) represent the simple and twisted interface models, respectively.
In Fig.~\ref{fig:dos} (b), the pDoS of Fe(d) exhibits well-defined exchange splitting representing ferromagnetic order, and no significant difference between Fig. ~\ref{fig:dos} (b), (e), and (h).
Because this behavior does not change, as shown in Fig. ~\ref{fig:dos}(e) and (h), the ferromagnetic properties are not degraded by the presence of graphene, which is consistent with recent experimental measurements \cite{naganuma2020perpendicular}.
As shown in Fig.~\ref{fig:dos}(c), the free-standing graphene has a characteristic $\mathrm{p}\pi$ band edge, the so-called \color{blue} Dirac core \color{black}, where pDoS becomes zero at $E_\mathrm{F}$.
For FePd(001)/Gr, as shown in Fig. ~\ref{fig:dos}(f) and (i), graphene has a hybrid state of $\mathrm{p}\pi$ and spin-polarized d-bands around $E_\mathrm{F}$.
This orbital hybridization originates from the chemical bonds between C and Fe, which leads to the induced magnetization of graphene, which is often observed in the graphene-ferromagnet interfaces \cite{dedkov2010electronic, abtew2013graphene}.

\begin{figure}
    \includegraphics[width=0.35\textwidth]{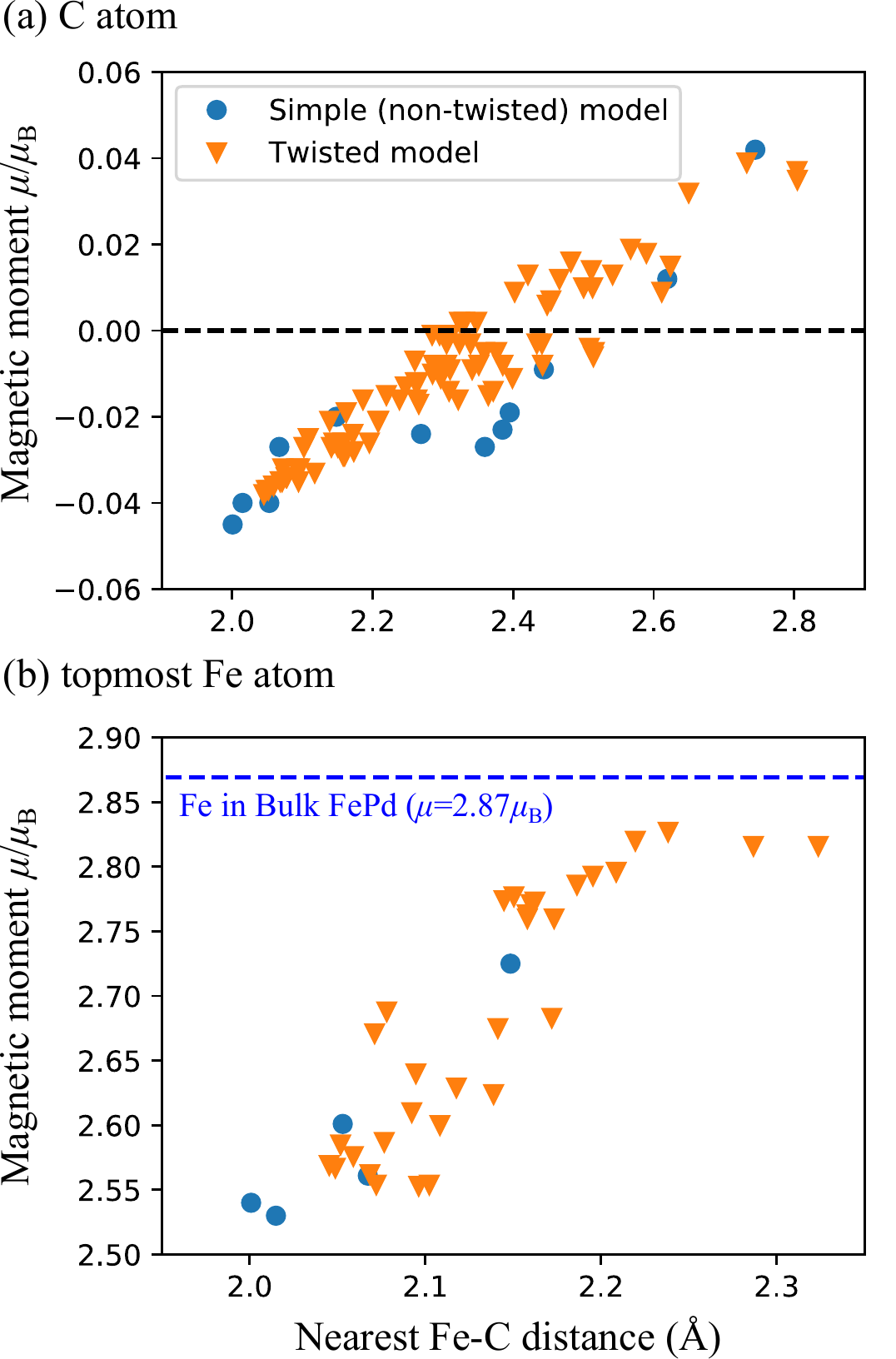}
    \caption{
        \label{fig:moment}
        Local magnetic moment on C atom (a) and topmost Fe atom (b) as a function of distance to nearest Fe (or C) atom.
        Both of the simple $S_{5,3}$ model (blue point) and twisted $T_{4,8,-5,2,-1,8}$ model (orange triangle) are also plotted.
        The blue line indicates that on the Fe atom in bulk FePd.
    }
\end{figure}

Figure~\ref{fig:moment}(a) shows the local magnetic moment of C atoms as a function of the distance to the nearest Fe atoms.
The simple model (blue points) and twisted model (orange triangles) indicate the same distribution.
The induced magnetic moment $\mu$ of C atoms far from Fe atoms is positive (the majority spin), which is the same polarization component of Fe(3d); this behavior results from charge transfer from the substrate to graphene.
In contrast, when the C atom nears the Fe atom, the magnetic moment gradually changes to the minority spin side. 
This is owing to the presence of minority spin d-states near $E_F$ [see Fig. ~\ref{fig:dos}(b)], where stronger hybridization increases the minority spin pDoS of $\mathrm{p}\pi$ band at $E_\mathrm{F}$.
The total magnetic moments of the graphene layer, because the contribution from each C atom is averaged, approaches a small value $|\mu| < 0.02\mu_\mathrm{B}$, which is below that of typical graphene on ferromagnetic metals \textit{ for example } Ni(111)/Gr of $\mu=0.05\mu_\mathrm{B} \sim 0.10\mu_\mathrm{B}$ \cite{dedkov2010electronic} \color{blue}.\color{black}
The local magnetic moment of the topmost Fe atoms is plotted in Fig. ~\ref{fig:moment}(b).
The decrease in the distance to the nearest C atom led to a decrease in the magnetic moment from that of a Fe atom in bulk FePd ($\mu \approx 2.87 \color{blue} \mu_\mathrm{B}$).
Similar behavior has been theoretically predicted for Ni(111)/Gr \cite{abtew2013graphene}.

\begin{figure}
    \centering
    \includegraphics[width=0.40\textwidth]{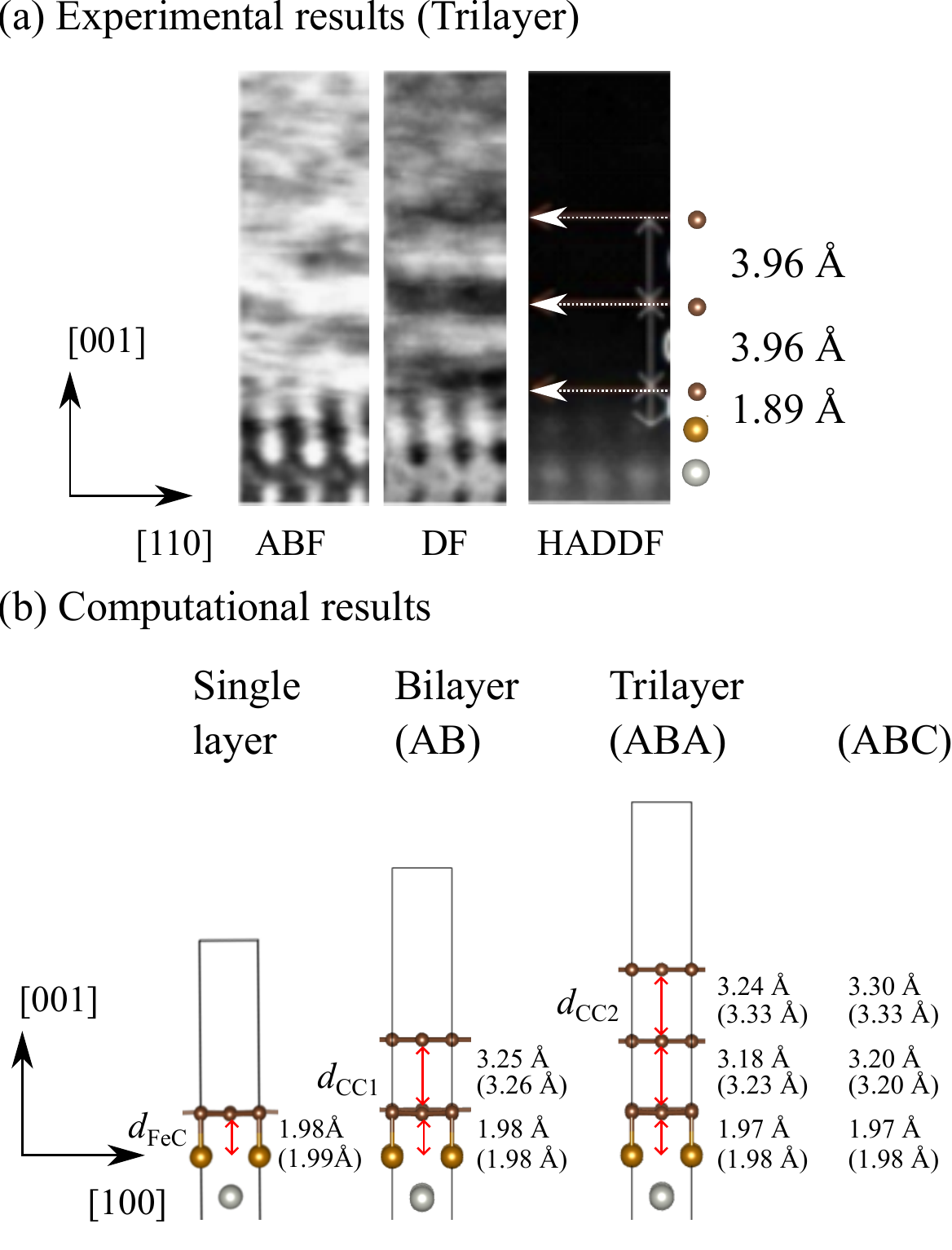}
    \caption{
        \label{fig:stem} 
        Comparison of the computational results and observation by the electron microscopy. 
        (a) STEM image of trilayer graphene on FePd interface.
        \color{blue}
        Detailed STEM images can be also seen in Ref~\cite{naganuma2022unveiling}
        \color{black}
        (b) Estimated interlayer distances of single-layer, AB-stacked bilayer, and ABA- and ABC-stacked trilayer graphene on FePd.
        \color{blue}
        The functionals of DFT-D2 and optB86b-vdW (the values in parentheses) are employed in this calculation.
    }
\end{figure}

Finally, figure~\ref{fig:stem}(a) shows the cross-sectional scanning tunneling electron microscopy (STEM) images of a sample in which three layers of graphene are formed on an epitaxial FePd film. Three complementary detection modes are presented: BF, ABF, and HAADF. From the BF and ABF-STEM images, it can be confirmed that the interatomic distance between the first layer of graphene and FePd is approximately $1.9~\text{\AA}$, whereas the upper layers of graphene are approximately $4~\text{\AA}$ thick. The HADDF-STEM image shows that Fe and Pd are alternately aligned in the out-of-plane direction, indicating a highly ordered $L1_0$ structure. 
The fact that the top surface layer is Fe is in agreement with the model shown in Fig. ~\ref{fig:slab}.
In Fig.~\ref{fig:stem}(b), we show the optimized interatomic distances of single-layer, bilayer, and trilayer graphene on a FePd slab.
\color{blue}
The distance between the lower graphene and Fe layers ($d_\mathrm{FeC}$) is close to that of the experimentally observed value.
The interlayer distances of the graphite-graphite layer ($d_\mathrm{CC1}$ and $d_\mathrm{CC2}$) is about $3.3~\text{\AA}$ (details are given in the supplementary material S2-S3) has discrepancy with experimental results. 
\color{black}

\section{\label{sec:summary} Summary}
This study investigated the atomic-scale structures, electronic and magnetic properties, and adsorption mechanism of the FePd(001)/Gr heterogeneous interface.
We proposed various atomic-scale models, including simple nontwisted and low-strain twisted interfaces.
The "simple model" is designed to consist of a small number of atoms, thereby saving computational costs. It provides an energetically (meta)stable structure; however, it has unrealistically large strain energy.
The later "twisted model" requires large supercells containing many atoms, but the strain can be greatly reduced.
We explored possible interface structures that are smaller than $9 \times 9$ supercell structures with lower strain (small MAE~$a_\mathrm{C}$).
The optimized structure had a binding energy of $E_\mathrm{B}=-0.22$~eV/atom in DFT-D2 ($E_\mathrm{B}=-0.19$~eV/atom in optB86b-vdW), which is comparable in order of magnitude to that of other well-understood metal/Gr interfaces.
The stability is governed by the strain of graphene and– C-Fe interatomic attraction.
Additionally, more stable structures are likely to exist in larger supercells (more than $10 \times 10$).
However, the present structure contains more than 300 atoms and larger supercell models are computationally difficult.
Further, because the strain energy $E_S$ in the present structure is already sufficiently small ($E_S \leq 0.02$~eV/atom), no significant $E_B$ improvements are expected even in the larger supercells.

The potential energy surface (PES) profiles shown in Fig. ~\ref{fig:shift} shows that the change in $E_\mathrm{B}$ is small ($\leq 5~\%$) for atomic-scale lateral shifts of graphene, which corresponds to the robustness of the absorption.
The flatness of PES also indicates suppression of the site-dependence of $E_\mathrm{B}$.
Typical lattice-matched interfaces (e.g., Ni(111)/Gr), owing to the fixed relative positions between the C and metal atoms, have sharp minimum points in the PES, resulting in characteristic adsorption sites.
For FePd(001)/Gr with a lattice symmetry mismatch, the relative positions of each Fe-C were uniformly distributed. 
As shown by the radial distribution function in Fig. ~\ref{fig:radial}, the interatomic attraction is averaged over the various configurations, which leads to a lack of site dependency in the PES.
Besides, the coexistence of chemisorption and physisorption distances results in intermediate bonding mechanisms.

For interfaces with small lattice constant mismatches (for example, Ir(111)/Gr), the \color{blue} moir\'e \color{black} structure and site-dependent attracting interaction originate from vertical buckling.
Such behavior in the symmetry-mismatched interface is non-trivial; however, in the case of FePd(001)Gr, our results predict the existence of \color{blue} moir\'e \color{black}-like buckling, as shown in Fig. ~\ref{fig:height}.
Their amplitudes ($\leq 0.5~\text{\AA}$) are much smaller than those of the small-lattice-constant mismatched interface; this is considered to be the result of averaging the inter-atomic attraction.
The DoS in Fig.~\ref{fig:dos} indicates that the macroscopic electronic and magnetic properties of FePd were not significantly changed by the presence of graphene, which is also in good agreement with the experimental results \cite{naganuma2020perpendicular}.
We expect FePd(001)/Gr to exhibit desirable properties for MTJ applications. Because the graphene coverage is sufficiently stable to protect Fe from oxidization without significantly degrading the macroscopic magnetic properties, the FePd(001)/Gr interface can avoid large lattice mismatches such as those exhibited by the FePd(001)/MgO interface.

In addition, comparing the DoS of the "simple model" and "twisted model" in Fig. ~\ref{fig:dos}, it is remarkable that the choice of the interface structure does not make a significant difference, especially near $E_\mathrm{F}$.
In a typical lattice-matched interface, 
the choice of adsorption site strongly influences their electronic properties.
However, in lattice symmetry-mismatched interfaces, site dependency is suppressed by the uniform averaging of Fe-C interactions.
This finding justifies the use of the simple model as a reasonable substitute for realistic twisted models in calculations of physical properties, which are mainly dominated by electronic and magnetic structures.
\color{blue}
(The Simple models are, due to the large strain,  inaccuracies for calculation related to structural energies and forces such as molecular dynamics and phonon properties.)
\color{black}
In the future, we expect that the simple models will become a useful testbed to reduce the computational costs of time-consuming calculations \textit{ that is }, electron transport properties \cite{egami2021calculation, ono2012first} and magnetocrystalline anisotropies \cite{blanco2019validity} of FePd(001)/Gr and any other $L1_0$ alloy-based interfaces.

\color{blue}
\section*{SUPPLEMENTARY MATERIAL}
See the supplementary material for optimized of atomic positions of the simple models (S1), and detailed computational results of bilayer (S2) and trilayer (S3) graphene on FePd(001) surface.
\color{black}

\begin{acknowledgments}

This work is supported by MEXT as "Program for Promoting Researches on the Supercomputer Fugaku" (Quantum-Theory-Based Multiscale Simulations
Toward the development of next-generation energy-saving semiconductor devices, JPMXP1020200205) and used the computational resources of super computer Fugaku provided by the RIKEN Center for Computational Science (Project ID: hp200122/hp150273).
T.O. acknowledge the financial support from JSPS KAKENHI (grant number:JP16H03865).
M.U., H.N., and T.O. acknowledges the JSPS Core-to-Core Program (grant number:JPJSCCA20160005).
This work is performed with the approval of the Photon Factory Program Advisory Committee (H.N.) (No. 2019S2-003).
The visualization of the atomic structures in this study is performed using VESTA software \cite{momma2011vesta} .
The computation is performed by the facilities of the Supercomputer Center, the Institute for Solid State Physics, the University of Tokyo, and also performed by Oakforest-PACS at JCAHPC through the Multidisciplinary Cooperative Research Program in CCS, the University of Tsukuba.
\end{acknowledgments}

\appendix

\nocite{*}

\bibliography{refs}

\begin{thebibliography}{34}%
\makeatletter
\providecommand \@ifxundefined [1]{%
 \@ifx{#1\undefined}
}%
\providecommand \@ifnum [1]{%
 \ifnum #1\expandafter \@firstoftwo
 \else \expandafter \@secondoftwo
 \fi
}%
\providecommand \@ifx [1]{%
 \ifx #1\expandafter \@firstoftwo
 \else \expandafter \@secondoftwo
 \fi
}%
\providecommand \natexlab [1]{#1}%
\providecommand \enquote  [1]{``#1''}%
\providecommand \bibnamefont  [1]{#1}%
\providecommand \bibfnamefont [1]{#1}%
\providecommand \citenamefont [1]{#1}%
\providecommand \href@noop [0]{\@secondoftwo}%
\providecommand \href [0]{\begingroup \@sanitize@url \@href}%
\providecommand \@href[1]{\@@startlink{#1}\@@href}%
\providecommand \@@href[1]{\endgroup#1\@@endlink}%
\providecommand \@sanitize@url [0]{\catcode `\\12\catcode `\$12\catcode
  `\&12\catcode `\#12\catcode `\^12\catcode `\_12\catcode `\%12\relax}%
\providecommand \@@startlink[1]{}%
\providecommand \@@endlink[0]{}%
\providecommand \url  [0]{\begingroup\@sanitize@url \@url }%
\providecommand \@url [1]{\endgroup\@href {#1}{\urlprefix }}%
\providecommand \urlprefix  [0]{URL }%
\providecommand \Eprint [0]{\href }%
\providecommand \doibase [0]{https://doi.org/}%
\providecommand \selectlanguage [0]{\@gobble}%
\providecommand \bibinfo  [0]{\@secondoftwo}%
\providecommand \bibfield  [0]{\@secondoftwo}%
\providecommand \translation [1]{[#1]}%
\providecommand \BibitemOpen [0]{}%
\providecommand \bibitemStop [0]{}%
\providecommand \bibitemNoStop [0]{.\EOS\space}%
\providecommand \EOS [0]{\spacefactor3000\relax}%
\providecommand \BibitemShut  [1]{\csname bibitem#1\endcsname}%
\let\auto@bib@innerbib\@empty
\bibitem [{\citenamefont {Naganuma}\ \emph {et~al.}(2015)\citenamefont
  {Naganuma}, \citenamefont {Kim}, \citenamefont {Kawada}, \citenamefont
  {Inami}, \citenamefont {Hatakeyama}, \citenamefont {Iihama}, \citenamefont
  {Nazrul~Islam}, \citenamefont {Oogane}, \citenamefont {Mizukami},\ and\
  \citenamefont {Ando}}]{naganuma2015electrical}%
  \BibitemOpen
  \bibfield  {author} {\bibinfo {author} {\bibfnamefont {H.}~\bibnamefont
  {Naganuma}}, \bibinfo {author} {\bibfnamefont {G.}~\bibnamefont {Kim}},
  \bibinfo {author} {\bibfnamefont {Y.}~\bibnamefont {Kawada}}, \bibinfo
  {author} {\bibfnamefont {N.}~\bibnamefont {Inami}}, \bibinfo {author}
  {\bibfnamefont {K.}~\bibnamefont {Hatakeyama}}, \bibinfo {author}
  {\bibfnamefont {S.}~\bibnamefont {Iihama}}, \bibinfo {author} {\bibfnamefont
  {K.~M.}\ \bibnamefont {Nazrul~Islam}}, \bibinfo {author} {\bibfnamefont
  {M.}~\bibnamefont {Oogane}}, \bibinfo {author} {\bibfnamefont
  {S.}~\bibnamefont {Mizukami}},\ and\ \bibinfo {author} {\bibfnamefont
  {Y.}~\bibnamefont {Ando}},\ }\bibfield  {title} {\enquote {\bibinfo {title}
  {{Electrical detection of millimeter-waves by magnetic tunnel junctions using
  perpendicular magnetized $L1_0$-FePd free layer}},}\ }\href
  {https://doi.org/10.1021/nl504114v} {\bibfield  {journal} {\bibinfo
  {journal} {Nano Lett.}\ }\textbf {\bibinfo {volume} {15}},\ \bibinfo {pages}
  {623--628} (\bibinfo {year} {2015})}\BibitemShut {NoStop}%
\bibitem [{\citenamefont {Zhang}\ \emph {et~al.}(2018)\citenamefont {Zhang},
  \citenamefont {Schliep}, \citenamefont {Wu}, \citenamefont {Quarterman},
  \citenamefont {Reifsnyder~Hickey}, \citenamefont {Lv}, \citenamefont {Chao},
  \citenamefont {Li}, \citenamefont {Chen}, \citenamefont {Zhao} \emph
  {et~al.}}]{zhang2018enhancement}%
  \BibitemOpen
  \bibfield  {author} {\bibinfo {author} {\bibfnamefont {D.-L.}\ \bibnamefont
  {Zhang}}, \bibinfo {author} {\bibfnamefont {K.~B.}\ \bibnamefont {Schliep}},
  \bibinfo {author} {\bibfnamefont {R.~J.}\ \bibnamefont {Wu}}, \bibinfo
  {author} {\bibfnamefont {P.}~\bibnamefont {Quarterman}}, \bibinfo {author}
  {\bibfnamefont {D.}~\bibnamefont {Reifsnyder~Hickey}}, \bibinfo {author}
  {\bibfnamefont {Y.}~\bibnamefont {Lv}}, \bibinfo {author} {\bibfnamefont
  {X.}~\bibnamefont {Chao}}, \bibinfo {author} {\bibfnamefont {H.}~\bibnamefont
  {Li}}, \bibinfo {author} {\bibfnamefont {J.-Y.}\ \bibnamefont {Chen}},
  \bibinfo {author} {\bibfnamefont {Z.}~\bibnamefont {Zhao}}, \emph {et~al.},\
  }\bibfield  {title} {\enquote {\bibinfo {title} {{Enhancement of tunneling
  magnetoresistance by inserting a diffusion barrier in L10-FePd perpendicular
  magnetic tunnel junctions}},}\ }\href {https://doi.org/10.1063/1.5019193}
  {\bibfield  {journal} {\bibinfo  {journal} {Appl. Phys. Lett.}\ }\textbf
  {\bibinfo {volume} {112}},\ \bibinfo {pages} {152401} (\bibinfo {year}
  {2018})}\BibitemShut {NoStop}%
\bibitem [{\citenamefont {Klemmer}\ \emph {et~al.}(1995)\citenamefont
  {Klemmer}, \citenamefont {Hoydick}, \citenamefont {Okumura}, \citenamefont
  {Zhang},\ and\ \citenamefont {Soffa}}]{klemmer1995magnetic}%
  \BibitemOpen
  \bibfield  {author} {\bibinfo {author} {\bibfnamefont {T.}~\bibnamefont
  {Klemmer}}, \bibinfo {author} {\bibfnamefont {D.}~\bibnamefont {Hoydick}},
  \bibinfo {author} {\bibfnamefont {H.}~\bibnamefont {Okumura}}, \bibinfo
  {author} {\bibfnamefont {B.}~\bibnamefont {Zhang}},\ and\ \bibinfo {author}
  {\bibfnamefont {W.}~\bibnamefont {Soffa}},\ }\bibfield  {title} {\enquote
  {\bibinfo {title} {{Magnetic hardening and coercivity mechanisms in $L1_0$
  ordered FePd ferromagnets}},}\ }\href
  {https://doi.org/10.1016/0956-716X(95)00413-P} {\bibfield  {journal}
  {\bibinfo  {journal} {Scr. Mater.}\ }\textbf {\bibinfo {volume} {33}},\
  \bibinfo {pages} {1793--1805} (\bibinfo {year} {1995})}\BibitemShut {NoStop}%
\bibitem [{\citenamefont {Shima}\ \emph {et~al.}(2004)\citenamefont {Shima},
  \citenamefont {Oikawa}, \citenamefont {Fujita}, \citenamefont {Fukamichi},
  \citenamefont {Ishida},\ and\ \citenamefont {Sakuma}}]{shima2004lattice}%
  \BibitemOpen
  \bibfield  {author} {\bibinfo {author} {\bibfnamefont {H.}~\bibnamefont
  {Shima}}, \bibinfo {author} {\bibfnamefont {K.}~\bibnamefont {Oikawa}},
  \bibinfo {author} {\bibfnamefont {A.}~\bibnamefont {Fujita}}, \bibinfo
  {author} {\bibfnamefont {K.}~\bibnamefont {Fukamichi}}, \bibinfo {author}
  {\bibfnamefont {K.}~\bibnamefont {Ishida}},\ and\ \bibinfo {author}
  {\bibfnamefont {A.}~\bibnamefont {Sakuma}},\ }\bibfield  {title} {\enquote
  {\bibinfo {title} {{Lattice axial ratio and large uniaxial magnetocrystalline
  anisotropy in $L{1}_{0}$-type FePd single crystals prepared under compressive
  stress}},}\ }\href {https://doi.org/10.1103/PhysRevB.70.224408} {\bibfield
  {journal} {\bibinfo  {journal} {Phys. Rev. B}\ }\textbf {\bibinfo {volume}
  {70}},\ \bibinfo {pages} {224408} (\bibinfo {year} {2004})}\BibitemShut
  {NoStop}%
\bibitem [{\citenamefont {Miyata}\ \emph {et~al.}(1990)\citenamefont {Miyata},
  \citenamefont {Asami}, \citenamefont {Mizushima},\ and\ \citenamefont
  {Sato}}]{miyata1990ferromagnetic}%
  \BibitemOpen
  \bibfield  {author} {\bibinfo {author} {\bibfnamefont {N.}~\bibnamefont
  {Miyata}}, \bibinfo {author} {\bibfnamefont {H.}~\bibnamefont {Asami}},
  \bibinfo {author} {\bibfnamefont {T.}~\bibnamefont {Mizushima}},\ and\
  \bibinfo {author} {\bibfnamefont {K.}~\bibnamefont {Sato}},\ }\bibfield
  {title} {\enquote {\bibinfo {title} {{Ferromagnetic Crystalline Anisotropy of
  Pd${}_{1-x}$ Fe${}_x$ Alloys. III. $0.38 \sim 0.5$, $L1_0$-Type Ordered
  Phase}},}\ }\href {https://doi.org/10.1143/JPSJ.59.1817} {\bibfield
  {journal} {\bibinfo  {journal} {J. Phys. Soc. Jpn.}\ }\textbf {\bibinfo
  {volume} {59}},\ \bibinfo {pages} {1817--1824} (\bibinfo {year}
  {1990})}\BibitemShut {NoStop}%
\bibitem [{\citenamefont {Naganuma}\ \emph {et~al.}(2020)\citenamefont
  {Naganuma}, \citenamefont {Zatko}, \citenamefont {Galbiati}, \citenamefont
  {Godel}, \citenamefont {Sander}, \citenamefont {Carr{\'e}t{\'e}ro},
  \citenamefont {Bezencenet}, \citenamefont {Reyren}, \citenamefont {Martin},
  \citenamefont {Dlubak} \emph {et~al.}}]{naganuma2020perpendicular}%
  \BibitemOpen
  \bibfield  {author} {\bibinfo {author} {\bibfnamefont {H.}~\bibnamefont
  {Naganuma}}, \bibinfo {author} {\bibfnamefont {V.}~\bibnamefont {Zatko}},
  \bibinfo {author} {\bibfnamefont {M.}~\bibnamefont {Galbiati}}, \bibinfo
  {author} {\bibfnamefont {F.}~\bibnamefont {Godel}}, \bibinfo {author}
  {\bibfnamefont {A.}~\bibnamefont {Sander}}, \bibinfo {author} {\bibfnamefont
  {C.}~\bibnamefont {Carr{\'e}t{\'e}ro}}, \bibinfo {author} {\bibfnamefont
  {O.}~\bibnamefont {Bezencenet}}, \bibinfo {author} {\bibfnamefont
  {N.}~\bibnamefont {Reyren}}, \bibinfo {author} {\bibfnamefont {M.-B.}\
  \bibnamefont {Martin}}, \bibinfo {author} {\bibfnamefont {B.}~\bibnamefont
  {Dlubak}}, \emph {et~al.},\ }\bibfield  {title} {\enquote {\bibinfo {title}
  {{A perpendicular graphene/ferromagnet electrode for spintronics}},}\ }\href
  {https://doi.org/10.1063/1.5143567} {\bibfield  {journal} {\bibinfo
  {journal} {Appl. Phys. Lett.}\ }\textbf {\bibinfo {volume} {116}},\ \bibinfo
  {pages} {173101} (\bibinfo {year} {2020})}\BibitemShut {NoStop}%
\bibitem [{\citenamefont {Mohri}\ \emph {et~al.}(2001)\citenamefont {Mohri},
  \citenamefont {Horiuchi}, \citenamefont {Uzawa}, \citenamefont {Ibaragi},
  \citenamefont {Igarashi},\ and\ \citenamefont {Abe}}]{mohri2001theoretical}%
  \BibitemOpen
  \bibfield  {author} {\bibinfo {author} {\bibfnamefont {T.}~\bibnamefont
  {Mohri}}, \bibinfo {author} {\bibfnamefont {T.}~\bibnamefont {Horiuchi}},
  \bibinfo {author} {\bibfnamefont {H.}~\bibnamefont {Uzawa}}, \bibinfo
  {author} {\bibfnamefont {M.}~\bibnamefont {Ibaragi}}, \bibinfo {author}
  {\bibfnamefont {M.}~\bibnamefont {Igarashi}},\ and\ \bibinfo {author}
  {\bibfnamefont {F.}~\bibnamefont {Abe}},\ }\bibfield  {title} {\enquote
  {\bibinfo {title} {{Theoretical investigation of $L1_0$-disorder phase
  equilibria in Fe--Pd alloy system}},}\ }\href
  {https://doi.org/10.1016/S0925-8388(00)01408-0} {\bibfield  {journal}
  {\bibinfo  {journal} {J. Alloys Compd.}\ }\textbf {\bibinfo {volume} {317}},\
  \bibinfo {pages} {13--18} (\bibinfo {year} {2001})}\BibitemShut {NoStop}%
\bibitem [{\citenamefont {Iihama}\ \emph {et~al.}(2014)\citenamefont {Iihama},
  \citenamefont {Sakuma}, \citenamefont {Naganuma}, \citenamefont {Oogane},
  \citenamefont {Miyazaki}, \citenamefont {Mizukami},\ and\ \citenamefont
  {Ando}}]{iihama2014low}%
  \BibitemOpen
  \bibfield  {author} {\bibinfo {author} {\bibfnamefont {S.}~\bibnamefont
  {Iihama}}, \bibinfo {author} {\bibfnamefont {A.}~\bibnamefont {Sakuma}},
  \bibinfo {author} {\bibfnamefont {H.}~\bibnamefont {Naganuma}}, \bibinfo
  {author} {\bibfnamefont {M.}~\bibnamefont {Oogane}}, \bibinfo {author}
  {\bibfnamefont {T.}~\bibnamefont {Miyazaki}}, \bibinfo {author}
  {\bibfnamefont {S.}~\bibnamefont {Mizukami}},\ and\ \bibinfo {author}
  {\bibfnamefont {Y.}~\bibnamefont {Ando}},\ }\bibfield  {title} {\enquote
  {\bibinfo {title} {{Low precessional damping observed for $L1_0$-ordered FePd
  epitaxial thin films with large perpendicular magnetic anisotropy}},}\ }\href
  {https://doi.org/10.1063/1.4897547} {\bibfield  {journal} {\bibinfo
  {journal} {Appl. Phys. Lett.}\ }\textbf {\bibinfo {volume} {105}},\ \bibinfo
  {pages} {142403} (\bibinfo {year} {2014})}\BibitemShut {NoStop}%
\bibitem [{\citenamefont {Kawai}\ \emph {et~al.}(2014)\citenamefont {Kawai},
  \citenamefont {Itabashi}, \citenamefont {Ohtake}, \citenamefont {Takeda},\
  and\ \citenamefont {Futamoto}}]{kawai2014gilbert}%
  \BibitemOpen
  \bibfield  {author} {\bibinfo {author} {\bibfnamefont {T.}~\bibnamefont
  {Kawai}}, \bibinfo {author} {\bibfnamefont {A.}~\bibnamefont {Itabashi}},
  \bibinfo {author} {\bibfnamefont {M.}~\bibnamefont {Ohtake}}, \bibinfo
  {author} {\bibfnamefont {S.}~\bibnamefont {Takeda}},\ and\ \bibinfo {author}
  {\bibfnamefont {M.}~\bibnamefont {Futamoto}},\ }\bibfield  {title} {\enquote
  {\bibinfo {title} {{Gilbert damping constant of FePd alloy thin films
  estimated by broadband ferromagnetic resonance}},}\ }in\ \href
  {https://doi.org/10.1051/epjconf/20147502002} {\emph {\bibinfo {booktitle}
  {{EPJ Web of Conferences}}}},\ Vol.~\bibinfo {volume} {75}\ (\bibinfo
  {organization} {EDP Sciences},\ \bibinfo {year} {2014})\ p.\ \bibinfo {pages}
  {02002}\BibitemShut {NoStop}%
\bibitem [{\citenamefont {Zharkov}\ \emph {et~al.}(2014)\citenamefont
  {Zharkov}, \citenamefont {Moiseenko}, \citenamefont {Altunin}, \citenamefont
  {Nikolaeva}, \citenamefont {Zhigalov},\ and\ \citenamefont
  {Myagkov}}]{zharkov2014study}%
  \BibitemOpen
  \bibfield  {author} {\bibinfo {author} {\bibfnamefont {S.}~\bibnamefont
  {Zharkov}}, \bibinfo {author} {\bibfnamefont {E.}~\bibnamefont {Moiseenko}},
  \bibinfo {author} {\bibfnamefont {R.}~\bibnamefont {Altunin}}, \bibinfo
  {author} {\bibfnamefont {N.}~\bibnamefont {Nikolaeva}}, \bibinfo {author}
  {\bibfnamefont {V.}~\bibnamefont {Zhigalov}},\ and\ \bibinfo {author}
  {\bibfnamefont {V.}~\bibnamefont {Myagkov}},\ }\bibfield  {title} {\enquote
  {\bibinfo {title} {{Study of solid-state reactions and order-disorder
  transitions in Pd/$\alpha$-Fe (001) thin films}},}\ }\href
  {https://doi.org/10.1134/S0021364014070145} {\bibfield  {journal} {\bibinfo
  {journal} {JETP Lett.}\ }\textbf {\bibinfo {volume} {99}},\ \bibinfo {pages}
  {405--409} (\bibinfo {year} {2014})}\BibitemShut {NoStop}%
\bibitem [{\citenamefont {Itabashi}\ \emph {et~al.}(2013)\citenamefont
  {Itabashi}, \citenamefont {Ohtake}, \citenamefont {Ouchi}, \citenamefont
  {Kirino},\ and\ \citenamefont {Futamoto}}]{itabashi2013preparation}%
  \BibitemOpen
  \bibfield  {author} {\bibinfo {author} {\bibfnamefont {A.}~\bibnamefont
  {Itabashi}}, \bibinfo {author} {\bibfnamefont {M.}~\bibnamefont {Ohtake}},
  \bibinfo {author} {\bibfnamefont {S.}~\bibnamefont {Ouchi}}, \bibinfo
  {author} {\bibfnamefont {F.}~\bibnamefont {Kirino}},\ and\ \bibinfo {author}
  {\bibfnamefont {M.}~\bibnamefont {Futamoto}},\ }\bibfield  {title} {\enquote
  {\bibinfo {title} {{Preparation of $L1_0$ ordered FePd, FePt, and CoPt thin
  films with flat surfaces on MgO (001) single-crystal substrates}},}\ }in\
  \href {https://doi.org/10.1051/epjconf/20134007001} {\emph {\bibinfo
  {booktitle} {{EPJ Web of Conferences}}}},\ Vol.~\bibinfo {volume} {40}\
  (\bibinfo {organization} {EDP Sciences},\ \bibinfo {year} {2013})\ p.\
  \bibinfo {pages} {07001}\BibitemShut {NoStop}%
\bibitem [{\citenamefont {Naganuma}\ \emph {et~al.}(2022)\citenamefont
  {Naganuma}, \citenamefont {Nishijima}, \citenamefont {Adachi}, \citenamefont
  {Uemoto}, \citenamefont {Shinya}, \citenamefont {Yasui}, \citenamefont
  {Morioka}, \citenamefont {Hirata}, \citenamefont {Godel}, \citenamefont
  {Martin} \emph {et~al.}}]{naganuma2022unveiling}%
  \BibitemOpen
  \bibfield  {author} {\bibinfo {author} {\bibfnamefont {H.}~\bibnamefont
  {Naganuma}}, \bibinfo {author} {\bibfnamefont {M.}~\bibnamefont {Nishijima}},
  \bibinfo {author} {\bibfnamefont {H.}~\bibnamefont {Adachi}}, \bibinfo
  {author} {\bibfnamefont {M.}~\bibnamefont {Uemoto}}, \bibinfo {author}
  {\bibfnamefont {H.}~\bibnamefont {Shinya}}, \bibinfo {author} {\bibfnamefont
  {S.}~\bibnamefont {Yasui}}, \bibinfo {author} {\bibfnamefont
  {H.}~\bibnamefont {Morioka}}, \bibinfo {author} {\bibfnamefont
  {A.}~\bibnamefont {Hirata}}, \bibinfo {author} {\bibfnamefont
  {F.}~\bibnamefont {Godel}}, \bibinfo {author} {\bibfnamefont {M.-B.}\
  \bibnamefont {Martin}}, \emph {et~al.},\ }\bibfield  {title} {\enquote
  {\bibinfo {title} {{Unveiling a Chemisorbed Crystallographically
  Heterogeneous Graphene/$L1_0$-FePd Interface with a Robust and Perpendicular
  Orbital Moment}},}\ }\href
  {https://doi.org/https://doi.org/10.1021/acsnano.1c09843} {\bibfield
  {journal} {\bibinfo  {journal} {ACS nano}\ }\textbf {\bibinfo {volume}
  {16}},\ \bibinfo {pages} {4139--4151} (\bibinfo {year} {2022})}\BibitemShut
  {NoStop}%
\bibitem [{\citenamefont {Kent}\ and\ \citenamefont
  {Worledge}(2015)}]{kent2015new}%
  \BibitemOpen
  \bibfield  {author} {\bibinfo {author} {\bibfnamefont {A.~D.}\ \bibnamefont
  {Kent}}\ and\ \bibinfo {author} {\bibfnamefont {D.~C.}\ \bibnamefont
  {Worledge}},\ }\bibfield  {title} {\enquote {\bibinfo {title} {{A new spin on
  magnetic memories}},}\ }\href {https://doi.org/10.1038/nnano.2015.24}
  {\bibfield  {journal} {\bibinfo  {journal} {Nat. Nanotechnol.}\ }\textbf
  {\bibinfo {volume} {10}},\ \bibinfo {pages} {187--191} (\bibinfo {year}
  {2015})}\BibitemShut {NoStop}%
\bibitem [{\citenamefont {Parkin}\ \emph {et~al.}(2004)\citenamefont {Parkin},
  \citenamefont {Kaiser}, \citenamefont {Panchula}, \citenamefont {Rice},
  \citenamefont {Hughes}, \citenamefont {Samant},\ and\ \citenamefont
  {Yang}}]{parkin2004giant}%
  \BibitemOpen
  \bibfield  {author} {\bibinfo {author} {\bibfnamefont {S.~S.}\ \bibnamefont
  {Parkin}}, \bibinfo {author} {\bibfnamefont {C.}~\bibnamefont {Kaiser}},
  \bibinfo {author} {\bibfnamefont {A.}~\bibnamefont {Panchula}}, \bibinfo
  {author} {\bibfnamefont {P.~M.}\ \bibnamefont {Rice}}, \bibinfo {author}
  {\bibfnamefont {B.}~\bibnamefont {Hughes}}, \bibinfo {author} {\bibfnamefont
  {M.}~\bibnamefont {Samant}},\ and\ \bibinfo {author} {\bibfnamefont {S.-H.}\
  \bibnamefont {Yang}},\ }\bibfield  {title} {\enquote {\bibinfo {title}
  {{Giant tunnelling magnetoresistance at room temperature with MgO (100)
  tunnel barriers}},}\ }\href {https://doi.org/10.1038/nmat1257} {\bibfield
  {journal} {\bibinfo  {journal} {{Nat. Mat.}}\ }\textbf {\bibinfo {volume}
  {3}},\ \bibinfo {pages} {862--867} (\bibinfo {year} {2004})}\BibitemShut
  {NoStop}%
\bibitem [{\citenamefont {Bertoni}\ \emph {et~al.}(2005)\citenamefont
  {Bertoni}, \citenamefont {Calmels}, \citenamefont {Altibelli},\ and\
  \citenamefont {Serin}}]{bertoni2005first}%
  \BibitemOpen
  \bibfield  {author} {\bibinfo {author} {\bibfnamefont {G.}~\bibnamefont
  {Bertoni}}, \bibinfo {author} {\bibfnamefont {L.}~\bibnamefont {Calmels}},
  \bibinfo {author} {\bibfnamefont {A.}~\bibnamefont {Altibelli}},\ and\
  \bibinfo {author} {\bibfnamefont {V.}~\bibnamefont {Serin}},\ }\bibfield
  {title} {\enquote {\bibinfo {title} {{First-principles calculation of the
  electronic structure and EELS spectra at the graphene/Ni (111) interface}},}\
  }\href {https://doi.org/10.1103/PhysRevB.71.075402} {\bibfield  {journal}
  {\bibinfo  {journal} {Phys. Rev. B}\ }\textbf {\bibinfo {volume} {71}},\
  \bibinfo {pages} {075402} (\bibinfo {year} {2005})}\BibitemShut {NoStop}%
\bibitem [{\citenamefont {Kozlov}, \citenamefont {Vines},\ and\ \citenamefont
  {G\"orling}(2012)}]{kozlov2012bonding}%
  \BibitemOpen
  \bibfield  {author} {\bibinfo {author} {\bibfnamefont {S.~M.}\ \bibnamefont
  {Kozlov}}, \bibinfo {author} {\bibfnamefont {F.}~\bibnamefont {Vines}},\ and\
  \bibinfo {author} {\bibfnamefont {A.}~\bibnamefont {G\"orling}},\ }\bibfield
  {title} {\enquote {\bibinfo {title} {{Bonding Mechanisms of Graphene on Metal
  Surfaces}},}\ }\href {https://doi.org/10.1021/jp210667f} {\bibfield
  {journal} {\bibinfo  {journal} {J. Phys. Chem. C}\ }\textbf {\bibinfo
  {volume} {116}},\ \bibinfo {pages} {7360--7366} (\bibinfo {year}
  {2012})}\BibitemShut {NoStop}%
\bibitem [{\citenamefont {Hamada}\ and\ \citenamefont
  {Otani}(2010)}]{hamada2010comparative}%
  \BibitemOpen
  \bibfield  {author} {\bibinfo {author} {\bibfnamefont {I.}~\bibnamefont
  {Hamada}}\ and\ \bibinfo {author} {\bibfnamefont {M.}~\bibnamefont {Otani}},\
  }\bibfield  {title} {\enquote {\bibinfo {title} {{Comparative van der Waals
  density-functional study of graphene on metal surfaces}},}\ }\href
  {https://doi.org/10.1103/PhysRevB.82.153412} {\bibfield  {journal} {\bibinfo
  {journal} {Phys. Rev. B}\ }\textbf {\bibinfo {volume} {82}},\ \bibinfo
  {pages} {153412} (\bibinfo {year} {2010})}\BibitemShut {NoStop}%
\bibitem [{\citenamefont {Varykhalov}\ \emph {et~al.}(2008)\citenamefont
  {Varykhalov}, \citenamefont {S\'anchez-Barriga}, \citenamefont {Shikin},
  \citenamefont {Biswas}, \citenamefont {Vescovo}, \citenamefont {Rybkin},
  \citenamefont {Marchenko},\ and\ \citenamefont
  {Rader}}]{varykhalov2000electronic}%
  \BibitemOpen
  \bibfield  {author} {\bibinfo {author} {\bibfnamefont {A.}~\bibnamefont
  {Varykhalov}}, \bibinfo {author} {\bibfnamefont {J.}~\bibnamefont
  {S\'anchez-Barriga}}, \bibinfo {author} {\bibfnamefont {A.~M.}\ \bibnamefont
  {Shikin}}, \bibinfo {author} {\bibfnamefont {C.}~\bibnamefont {Biswas}},
  \bibinfo {author} {\bibfnamefont {E.}~\bibnamefont {Vescovo}}, \bibinfo
  {author} {\bibfnamefont {A.}~\bibnamefont {Rybkin}}, \bibinfo {author}
  {\bibfnamefont {D.}~\bibnamefont {Marchenko}},\ and\ \bibinfo {author}
  {\bibfnamefont {O.}~\bibnamefont {Rader}},\ }\bibfield  {title} {\enquote
  {\bibinfo {title} {{Electronic and Magnetic Properties of Quasifreestanding
  Graphene on Ni}},}\ }\href {https://doi.org/10.1103/PhysRevLett.101.157601}
  {\bibfield  {journal} {\bibinfo  {journal} {Phys. Rev. Lett.}\ }\textbf
  {\bibinfo {volume} {101}},\ \bibinfo {pages} {157601} (\bibinfo {year}
  {2008})}\BibitemShut {NoStop}%
\bibitem [{\citenamefont {Abtew}\ \emph {et~al.}(2013)\citenamefont {Abtew},
  \citenamefont {Shih}, \citenamefont {Banerjee},\ and\ \citenamefont
  {Zhang}}]{abtew2013graphene}%
  \BibitemOpen
  \bibfield  {author} {\bibinfo {author} {\bibfnamefont {T.}~\bibnamefont
  {Abtew}}, \bibinfo {author} {\bibfnamefont {B.-C.}\ \bibnamefont {Shih}},
  \bibinfo {author} {\bibfnamefont {S.}~\bibnamefont {Banerjee}},\ and\
  \bibinfo {author} {\bibfnamefont {P.}~\bibnamefont {Zhang}},\ }\bibfield
  {title} {\enquote {\bibinfo {title} {{Graphene--ferromagnet interfaces:
  hybridization, magnetization and charge transfer}},}\ }\href
  {https://doi.org/10.1039/c2nr32972g} {\bibfield  {journal} {\bibinfo
  {journal} {Nanoscale}\ }\textbf {\bibinfo {volume} {5}},\ \bibinfo {pages}
  {1902--1909} (\bibinfo {year} {2013})}\BibitemShut {NoStop}%
\bibitem [{\citenamefont {Mittendorfer}\ \emph {et~al.}(2011)\citenamefont
  {Mittendorfer}, \citenamefont {Garhofer}, \citenamefont {Redinger},
  \citenamefont {Klime\ifmmode~\check{s}\else \v{s}\fi{}}, \citenamefont
  {Harl},\ and\ \citenamefont {Kresse}}]{mittendorfer2011graphene}%
  \BibitemOpen
  \bibfield  {author} {\bibinfo {author} {\bibfnamefont {F.}~\bibnamefont
  {Mittendorfer}}, \bibinfo {author} {\bibfnamefont {A.}~\bibnamefont
  {Garhofer}}, \bibinfo {author} {\bibfnamefont {J.}~\bibnamefont {Redinger}},
  \bibinfo {author} {\bibfnamefont {J.}~\bibnamefont
  {Klime\ifmmode~\check{s}\else \v{s}\fi{}}}, \bibinfo {author} {\bibfnamefont
  {J.}~\bibnamefont {Harl}},\ and\ \bibinfo {author} {\bibfnamefont
  {G.}~\bibnamefont {Kresse}},\ }\bibfield  {title} {\enquote {\bibinfo {title}
  {{Graphene on Ni(111): Strong interaction and weak adsorption}},}\ }\href
  {https://doi.org/10.1103/PhysRevB.84.201401} {\bibfield  {journal} {\bibinfo
  {journal} {Phys. Rev. B}\ }\textbf {\bibinfo {volume} {84}},\ \bibinfo
  {pages} {201401} (\bibinfo {year} {2011})}\BibitemShut {NoStop}%
\bibitem [{\citenamefont {Batzill}(2012)}]{Batzill2012}%
  \BibitemOpen
  \bibfield  {author} {\bibinfo {author} {\bibfnamefont {M.}~\bibnamefont
  {Batzill}},\ }\bibfield  {title} {\enquote {\bibinfo {title} {{The surface
  science of graphene: Metal interfaces, CVD synthesis, nanoribbons, chemical
  modifications, and defects}},}\ }\href
  {https://doi.org/https://doi.org/10.1016/j.surfrep.2011.12.001} {\bibfield
  {journal} {\bibinfo  {journal} {Surf. Sci. Rep.}\ }\textbf {\bibinfo {volume}
  {67}},\ \bibinfo {pages} {83--115} (\bibinfo {year} {2012})}\BibitemShut
  {NoStop}%
\bibitem [{\citenamefont {Wintterlin}\ and\ \citenamefont
  {Bocquet}(2009)}]{wintterlin2009graphene}%
  \BibitemOpen
  \bibfield  {author} {\bibinfo {author} {\bibfnamefont {J.}~\bibnamefont
  {Wintterlin}}\ and\ \bibinfo {author} {\bibfnamefont {M.-L.}\ \bibnamefont
  {Bocquet}},\ }\bibfield  {title} {\enquote {\bibinfo {title} {{Graphene on
  metal surfaces}},}\ }\href
  {https://doi.org/https://doi.org/10.1016/j.susc.2008.08.037} {\bibfield
  {journal} {\bibinfo  {journal} {Surf. Sci.}\ }\textbf {\bibinfo {volume}
  {603}},\ \bibinfo {pages} {1841--1852} (\bibinfo {year} {2009})}\BibitemShut
  {NoStop}%
\bibitem [{\citenamefont {Hasegawa}\ \emph {et~al.}(2013)\citenamefont
  {Hasegawa}, \citenamefont {Nishidate}, \citenamefont {Hosokai},\ and\
  \citenamefont {Yoshimoto}}]{hasegawa2013electronic}%
  \BibitemOpen
  \bibfield  {author} {\bibinfo {author} {\bibfnamefont {M.}~\bibnamefont
  {Hasegawa}}, \bibinfo {author} {\bibfnamefont {K.}~\bibnamefont {Nishidate}},
  \bibinfo {author} {\bibfnamefont {T.}~\bibnamefont {Hosokai}},\ and\ \bibinfo
  {author} {\bibfnamefont {N.}~\bibnamefont {Yoshimoto}},\ }\bibfield  {title}
  {\enquote {\bibinfo {title} {Electronic-structure modification of graphene on
  ni (111) surface by the intercalation of a noble metal},}\ }\href
  {https://doi.org/10.1103/PhysRevB.87.085439} {\bibfield  {journal} {\bibinfo
  {journal} {Phys. Rev. B}\ }\textbf {\bibinfo {volume} {87}},\ \bibinfo
  {pages} {085439} (\bibinfo {year} {2013})}\BibitemShut {NoStop}%
\bibitem [{\citenamefont {Bl{\"o}chl}(1994)}]{blochl1994projector}%
  \BibitemOpen
  \bibfield  {author} {\bibinfo {author} {\bibfnamefont {P.~E.}\ \bibnamefont
  {Bl{\"o}chl}},\ }\bibfield  {title} {\enquote {\bibinfo {title} {{Projector
  augmented-wave method}},}\ }\href {https://doi.org/10.1103/PhysRevB.50.17953}
  {\bibfield  {journal} {\bibinfo  {journal} {Phys. Rev. B}\ }\textbf {\bibinfo
  {volume} {50}},\ \bibinfo {pages} {17953} (\bibinfo {year}
  {1994})}\BibitemShut {NoStop}%
\bibitem [{\citenamefont {Perdew}, \citenamefont {Burke},\ and\ \citenamefont
  {Ernzerhof}(1996)}]{perdew1996generalized}%
  \BibitemOpen
  \bibfield  {author} {\bibinfo {author} {\bibfnamefont {J.~P.}\ \bibnamefont
  {Perdew}}, \bibinfo {author} {\bibfnamefont {K.}~\bibnamefont {Burke}},\ and\
  \bibinfo {author} {\bibfnamefont {M.}~\bibnamefont {Ernzerhof}},\ }\bibfield
  {title} {\enquote {\bibinfo {title} {{Generalized gradient approximation made
  simple}},}\ }\href {https://doi.org/10.1103/PhysRevLett.77.3865} {\bibfield
  {journal} {\bibinfo  {journal} {Phys. Rev. Lett.}\ }\textbf {\bibinfo
  {volume} {77}},\ \bibinfo {pages} {3865} (\bibinfo {year}
  {1996})}\BibitemShut {NoStop}%
\bibitem [{\citenamefont {Grimme}(2006)}]{grimme2006semiempirical}%
  \BibitemOpen
  \bibfield  {author} {\bibinfo {author} {\bibfnamefont {S.}~\bibnamefont
  {Grimme}},\ }\bibfield  {title} {\enquote {\bibinfo {title} {{Semiempirical
  GGA-type density functional constructed with a long-range dispersion
  correction}},}\ }\href {https://doi.org/10.1002/jcc.20495} {\bibfield
  {journal} {\bibinfo  {journal} {J. Comp. Chem.}\ }\textbf {\bibinfo {volume}
  {27}},\ \bibinfo {pages} {1787--1799} (\bibinfo {year} {2006})}\BibitemShut
  {NoStop}%
\bibitem [{\citenamefont {Klime{\v{s}}}, \citenamefont {Bowler},\ and\
  \citenamefont {Michaelides}(2009)}]{klimes2009}%
  \BibitemOpen
  \bibfield  {author} {\bibinfo {author} {\bibfnamefont {J.}~\bibnamefont
  {Klime{\v{s}}}}, \bibinfo {author} {\bibfnamefont {D.~R.}\ \bibnamefont
  {Bowler}},\ and\ \bibinfo {author} {\bibfnamefont {A.}~\bibnamefont
  {Michaelides}},\ }\bibfield  {title} {\enquote {\bibinfo {title} {{Chemical
  accuracy for the van der Waals density functional}},}\ }\href
  {https://doi.org/10.1088/0953-8984/22/2/022201} {\bibfield  {journal}
  {\bibinfo  {journal} {J. Phys. Condens. Matter}\ }\textbf {\bibinfo {volume}
  {22}},\ \bibinfo {pages} {022201} (\bibinfo {year} {2009})}\BibitemShut
  {NoStop}%
\bibitem [{\citenamefont {Klime\ifmmode~\check{s}\else \v{s}\fi{}},
  \citenamefont {Bowler},\ and\ \citenamefont
  {Michaelides}(2011)}]{klimes2011}%
  \BibitemOpen
  \bibfield  {author} {\bibinfo {author} {\bibfnamefont {J.~c.~v.}\
  \bibnamefont {Klime\ifmmode~\check{s}\else \v{s}\fi{}}}, \bibinfo {author}
  {\bibfnamefont {D.~R.}\ \bibnamefont {Bowler}},\ and\ \bibinfo {author}
  {\bibfnamefont {A.}~\bibnamefont {Michaelides}},\ }\bibfield  {title}
  {\enquote {\bibinfo {title} {{Van der Waals density functionals applied to
  solids}},}\ }\href {https://doi.org/10.1103/PhysRevB.83.195131} {\bibfield
  {journal} {\bibinfo  {journal} {Phys. Rev. B}\ }\textbf {\bibinfo {volume}
  {83}},\ \bibinfo {pages} {195131} (\bibinfo {year} {2011})}\BibitemShut
  {NoStop}%
\bibitem [{\citenamefont {Piquemal-Banci}\ \emph {et~al.}(2020)\citenamefont
  {Piquemal-Banci}, \citenamefont {Galceran}, \citenamefont {Dubois},
  \citenamefont {Zatko}, \citenamefont {Galbiati}, \citenamefont {Godel},
  \citenamefont {Martin}, \citenamefont {Weatherup}, \citenamefont {Petroff},
  \citenamefont {Fert} \emph {et~al.}}]{piquemal2020spin}%
  \BibitemOpen
  \bibfield  {author} {\bibinfo {author} {\bibfnamefont {M.}~\bibnamefont
  {Piquemal-Banci}}, \bibinfo {author} {\bibfnamefont {R.}~\bibnamefont
  {Galceran}}, \bibinfo {author} {\bibfnamefont {S.~M.-M.}\ \bibnamefont
  {Dubois}}, \bibinfo {author} {\bibfnamefont {V.}~\bibnamefont {Zatko}},
  \bibinfo {author} {\bibfnamefont {M.}~\bibnamefont {Galbiati}}, \bibinfo
  {author} {\bibfnamefont {F.}~\bibnamefont {Godel}}, \bibinfo {author}
  {\bibfnamefont {M.-B.}\ \bibnamefont {Martin}}, \bibinfo {author}
  {\bibfnamefont {R.~S.}\ \bibnamefont {Weatherup}}, \bibinfo {author}
  {\bibfnamefont {F.}~\bibnamefont {Petroff}}, \bibinfo {author} {\bibfnamefont
  {A.}~\bibnamefont {Fert}}, \emph {et~al.},\ }\bibfield  {title} {\enquote
  {\bibinfo {title} {{Spin filtering by proximity effects at hybridized
  interfaces in spin-valves with 2D graphene barriers}},}\ }\href
  {https://doi.org/10.1038/s41467-020-19420-6} {\bibfield  {journal} {\bibinfo
  {journal} {Nat. Commun.}\ }\textbf {\bibinfo {volume} {11}},\ \bibinfo
  {pages} {1--9} (\bibinfo {year} {2020})}\BibitemShut {NoStop}%
\bibitem [{\citenamefont {Dedkov}\ and\ \citenamefont
  {Fonin}(2010)}]{dedkov2010electronic}%
  \BibitemOpen
  \bibfield  {author} {\bibinfo {author} {\bibfnamefont {Y.~S.}\ \bibnamefont
  {Dedkov}}\ and\ \bibinfo {author} {\bibfnamefont {M.}~\bibnamefont {Fonin}},\
  }\bibfield  {title} {\enquote {\bibinfo {title} {{Electronic and magnetic
  properties of the graphene--ferromagnet interface}},}\ }\href
  {https://doi.org/https://doi.org/10.1088/1367-2630/12/12/125004} {\bibfield
  {journal} {\bibinfo  {journal} {New J. Phys.}\ }\textbf {\bibinfo {volume}
  {12}},\ \bibinfo {pages} {125004} (\bibinfo {year} {2010})}\BibitemShut
  {NoStop}%
\bibitem [{\citenamefont {Egami}, \citenamefont {Tsukamoto},\ and\
  \citenamefont {Ono}(2021)}]{egami2021calculation}%
  \BibitemOpen
  \bibfield  {author} {\bibinfo {author} {\bibfnamefont {Y.}~\bibnamefont
  {Egami}}, \bibinfo {author} {\bibfnamefont {S.}~\bibnamefont {Tsukamoto}},\
  and\ \bibinfo {author} {\bibfnamefont {T.}~\bibnamefont {Ono}},\ }\bibfield
  {title} {\enquote {\bibinfo {title} {{Calculation of the Green's function in
  the scattering region for first-principles electron-transport
  simulations}},}\ }\href
  {https://doi.org/https://doi.org/10.1103/PhysRevResearch.3.013038} {\bibfield
   {journal} {\bibinfo  {journal} {Phys. Rev. Research}\ }\textbf {\bibinfo
  {volume} {3}},\ \bibinfo {pages} {013038} (\bibinfo {year}
  {2021})}\BibitemShut {NoStop}%
\bibitem [{\citenamefont {Ono}, \citenamefont {Egami},\ and\ \citenamefont
  {Hirose}(2012)}]{ono2012first}%
  \BibitemOpen
  \bibfield  {author} {\bibinfo {author} {\bibfnamefont {T.}~\bibnamefont
  {Ono}}, \bibinfo {author} {\bibfnamefont {Y.}~\bibnamefont {Egami}},\ and\
  \bibinfo {author} {\bibfnamefont {K.}~\bibnamefont {Hirose}},\ }\bibfield
  {title} {\enquote {\bibinfo {title} {First-principles transport calculation
  method based on real-space finite-difference nonequilibrium green's function
  scheme},}\ }\href
  {https://doi.org/https://doi.org/10.1103/PhysRevB.86.195406} {\bibfield
  {journal} {\bibinfo  {journal} {Phys. Rev. B}\ }\textbf {\bibinfo {volume}
  {86}},\ \bibinfo {pages} {195406} (\bibinfo {year} {2012})}\BibitemShut
  {NoStop}%
\bibitem [{\citenamefont {Blanco-Rey}, \citenamefont {Cerd{\'{a}}},\ and\
  \citenamefont {Arnau}(2019)}]{blanco2019validity}%
  \BibitemOpen
  \bibfield  {author} {\bibinfo {author} {\bibfnamefont {M.}~\bibnamefont
  {Blanco-Rey}}, \bibinfo {author} {\bibfnamefont {J.~I.}\ \bibnamefont
  {Cerd{\'{a}}}},\ and\ \bibinfo {author} {\bibfnamefont {A.}~\bibnamefont
  {Arnau}},\ }\bibfield  {title} {\enquote {\bibinfo {title} {Validity of
  perturbative methods to treat the spin{\textendash}orbit interaction:
  application to magnetocrystalline anisotropy},}\ }\href
  {https://doi.org/10.1088/1367-2630/ab3060} {\bibfield  {journal} {\bibinfo
  {journal} {New J. Phys.}\ }\textbf {\bibinfo {volume} {21}},\ \bibinfo
  {pages} {073054} (\bibinfo {year} {2019})}\BibitemShut {NoStop}%
\bibitem [{\citenamefont {Momma}\ and\ \citenamefont
  {Izumi}(2011)}]{momma2011vesta}%
  \BibitemOpen
  \bibfield  {author} {\bibinfo {author} {\bibfnamefont {K.}~\bibnamefont
  {Momma}}\ and\ \bibinfo {author} {\bibfnamefont {F.}~\bibnamefont {Izumi}},\
  }\bibfield  {title} {\enquote {\bibinfo {title} {{VESTA 3 for
  three-dimensional visualization of crystal, volumetric and morphology
  data}},}\ }\href {https://doi.org/10.1107/S0021889811038970} {\bibfield
  {journal} {\bibinfo  {journal} {J. Appl. Crystallogr.}\ }\textbf {\bibinfo
  {volume} {44}},\ \bibinfo {pages} {1272--1276} (\bibinfo {year}
  {2011})}\BibitemShut {NoStop}%
\end{thebibliography}%

\end{document}



\title{
Supporting Information of \\ "Density functional study of twisted graphene $L1_0$-FePd heterogeneous interface"
}

\author{Mitsuharu Uemoto}
\email{uemoto@eedept.kobe-u.ac.jp}
\affiliation{Department of Electrical and Electronic Engineering, Graduate School of Engineering, Kobe University, 1-1 Rokkodai-cho, Nada-ku, Kobe 651-8501, Japan}

\author{Hayato Adachi}
\affiliation{Department of Electrical and Electronic Engineering, Graduate School of Engineering, Kobe University, 1-1 Rokkodai-cho, Nada-ku, Kobe 651-8501, Japan}

\author{Hiroshi Naganuma}
\affiliation{Center for Innovative Integrated Electronics Systems (CIES), Tohoku University, 468-1 Aramaki Aza Aoba, Aoba, Sendai, Miyagi, 980-8572, Japan}
\affiliation{Center for Spintronics Integrated Systems (CSIS), Tohoku University, 2-2-1 Katahira Aoba, Sendai, Miyagi 980-8577 Japan}
\affiliation{Center for Spintronics Research Network (CSRN), Tohoku University, 2-1-1 Katahira, Aoba, Sendai, Miyagi 980-8577 Japan}
\affiliation{Graduate School of Engineering, Tohoku University, 6-6-05, Aoba, Aoba-ku, Sendai, Miyagi, 980-8579, Japan}

\author{Tomoya Ono}
\affiliation{Department of Electrical and Electronic Engineering, Graduate School of Engineering, Kobe University, 1-1 Rokkodai-cho, Nada-ku, Kobe 651-8501, Japan}

\date{\today}

\maketitle

\clearpage
\subsection*{S1. Optimized of atomic positions of the simple models}
\label{sec:structure}

\begin{figure}[h]
    \centering
    \includegraphics[width=90mm]{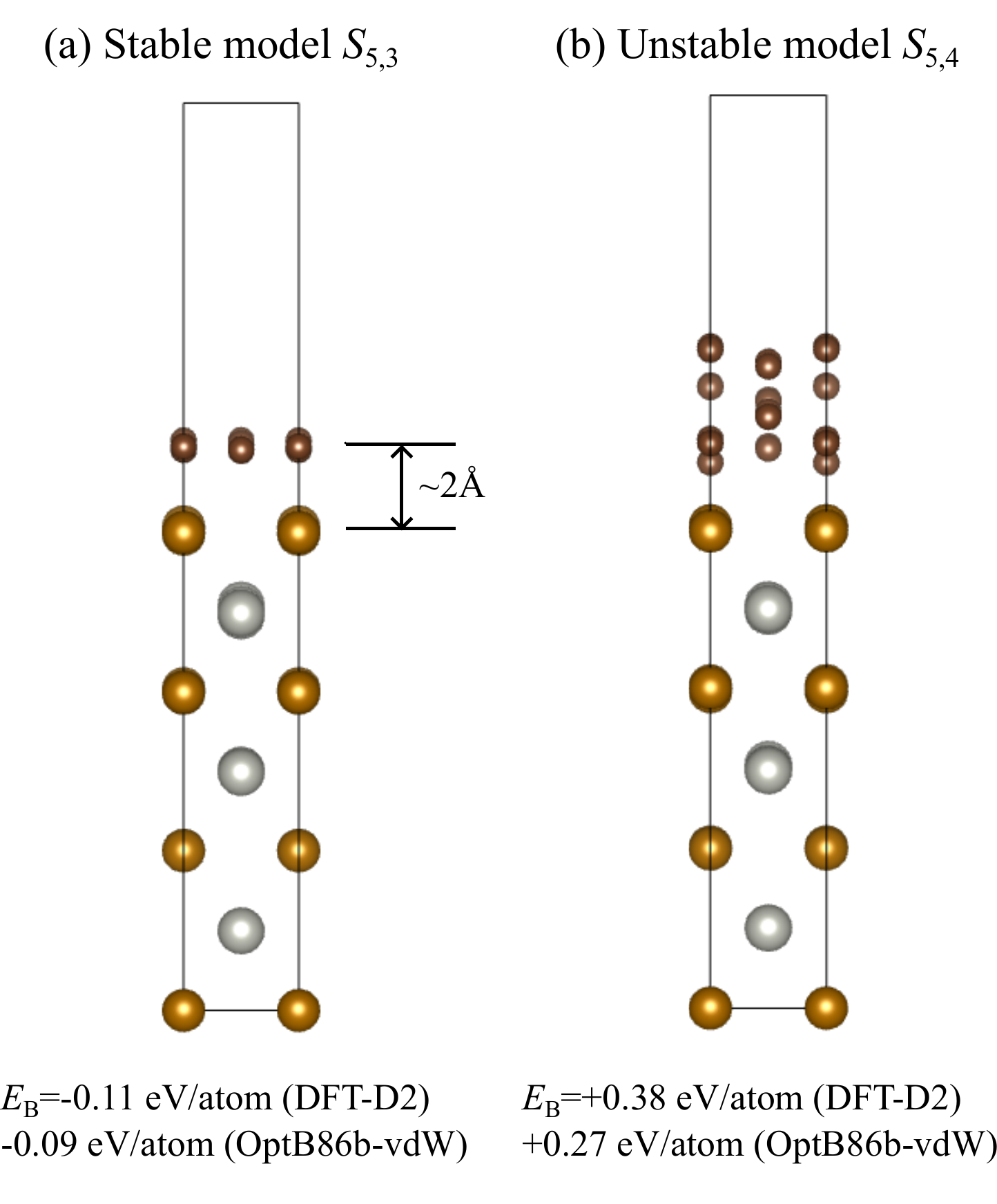}
    \caption{
         Optimized atomic position of stable $S_{5, 3}$ model (a) and unstable $S_{5, 4}$ model (b).
    }
    \label{fig:unstable}
\end{figure}

Figure.~\ref{fig:unstable} shows the atomic position of the typical stable and unstable interface models.
As examples, we pick two simple (non-twisted) models of $S_{5,3}$ and $S_{5,4}$.
The stable ($S_{5,3}$) model in Fig.~\ref{fig:unstable}(a), which has negative binding energy of $E_\mathrm{B} < -0.09$~eV/atom, exhibits the planar graphene layer; the interlayer distance is approximately $2~\text{\AA}$.
On the other hand, for the typical unstable ($S_{5,4}$) model  in Fig.~\ref{fig:unstable}(b), significant compressive or tensile strain leads to the destruction of flat graphene structures; such unstable structure has a large positive energy of $E_B > +0.27~\mathrm{eV/atom}$ and is physically unrealistic.

\clearpage
\subsection*{S2. Detail of bilayer graphene on FePd(001) surface}
\label{sec:bilayer}

\begin{figure}[h]
    \centering
    \includegraphics[width=140mm]{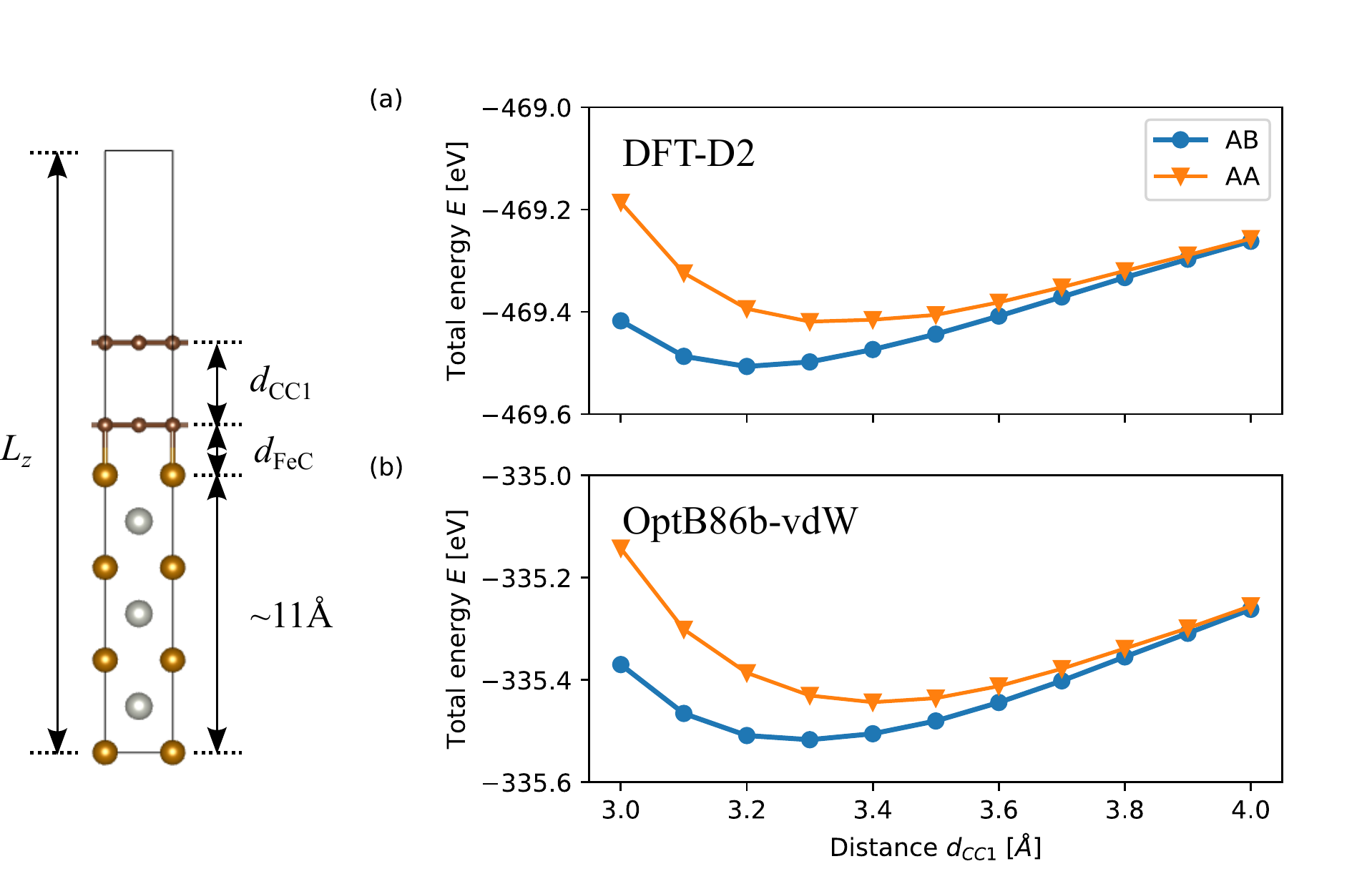}
    \caption{
        Total energy as a function of the graphene-graphene distance ($d_\mathrm{CC1}$) of AB- and AA-stacked bilayer graphenes on FePd(001) surface.
        Two different functionals of DFT-D2 (a) and OptB86b-vdW (b) are considered for comparison.
    }
    \label{fig:fepdgr2dist}
\end{figure}

\begin{table}[h]
\caption{
Optimized interlayer distances of AB-stacked bilayer graphene on FePd(001).
The functionals of DFT-D2 and OptB86b-vdW (the values in parentheses) are tested.
}

\label{tbl:bilayer}
\begin{tabular}{ccrr}
\toprule
Stacking order & Supercell size & \multicolumn{2}{c}{Interlayer distance} \\
 &  $L_z$~(\AA) & $d_\mathrm{FeC}$~(\AA) & $d_\mathrm{CC1}$~(\AA)  \\
\midrule
 AB & 23.1 & 1.98~(1.98) & 3.25~(3.26) \\
    & 24.1 & 1.98~(1.98) & 3.25~(3.26) \\
\bottomrule
\end{tabular}
\end{table}

To analyze the structure of bilayer graphene on $L1_0$-FePd(001), we propose an atomic structure model illustrated in Fig.~\ref{fig:fepdgr2dist}.
The supercell consists of seven atomic layers of FePd  (thickness about 11~\AA) and two graphene layers.
We assume the Fe/graphene interface of simple (non-twisted) $S_{5,3}$ model and consider the AB- and AA-stacking order for the graphene layers.
$L_z$ is the size of the supercell, and $d_\mathrm{FeC}$ and $d_\mathrm{CC1}$ are the interlayer distances.
The detailed computational conditions are the same as in the Manuscript.

Figure~\ref{fig:fepdgr2dist}(a)-(b) shows the total energy dependence on $d_\mathrm{CC1}$ without the structural relaxation (the calculations are restricted to fixed atomic positions).
For this calculation, we assume $L_z \approx 24~\text{\AA}$ and $d_\mathrm{FeC} \approx 2~\text{\AA}$.
We use two functionals, DFT-D2 and OptB86b-vdW, for comparison.
In both cases, the AB stacked bilayer with $d_\mathrm{CC1}=3.2 \sim 3.4~\text{\AA}$ is energetically the stable.

Next, structural relaxations are performed for the atomic positions.
The optimized distance between the atomic layers are shown in Table~\ref{tbl:bilayer}; two different $L_z$ are tested for convergence.
The obtained $d_\mathrm{CC1}$ values are slightly less than the well-known first-principles distance of $3.3~\text{\AA}$.

\clearpage
\subsection*{S3. Detail of trilayer graphene on FePd(001) surface}
\label{sec:trilayer}

\begin{figure}[h]
    \centering
    \includegraphics[width=120mm]{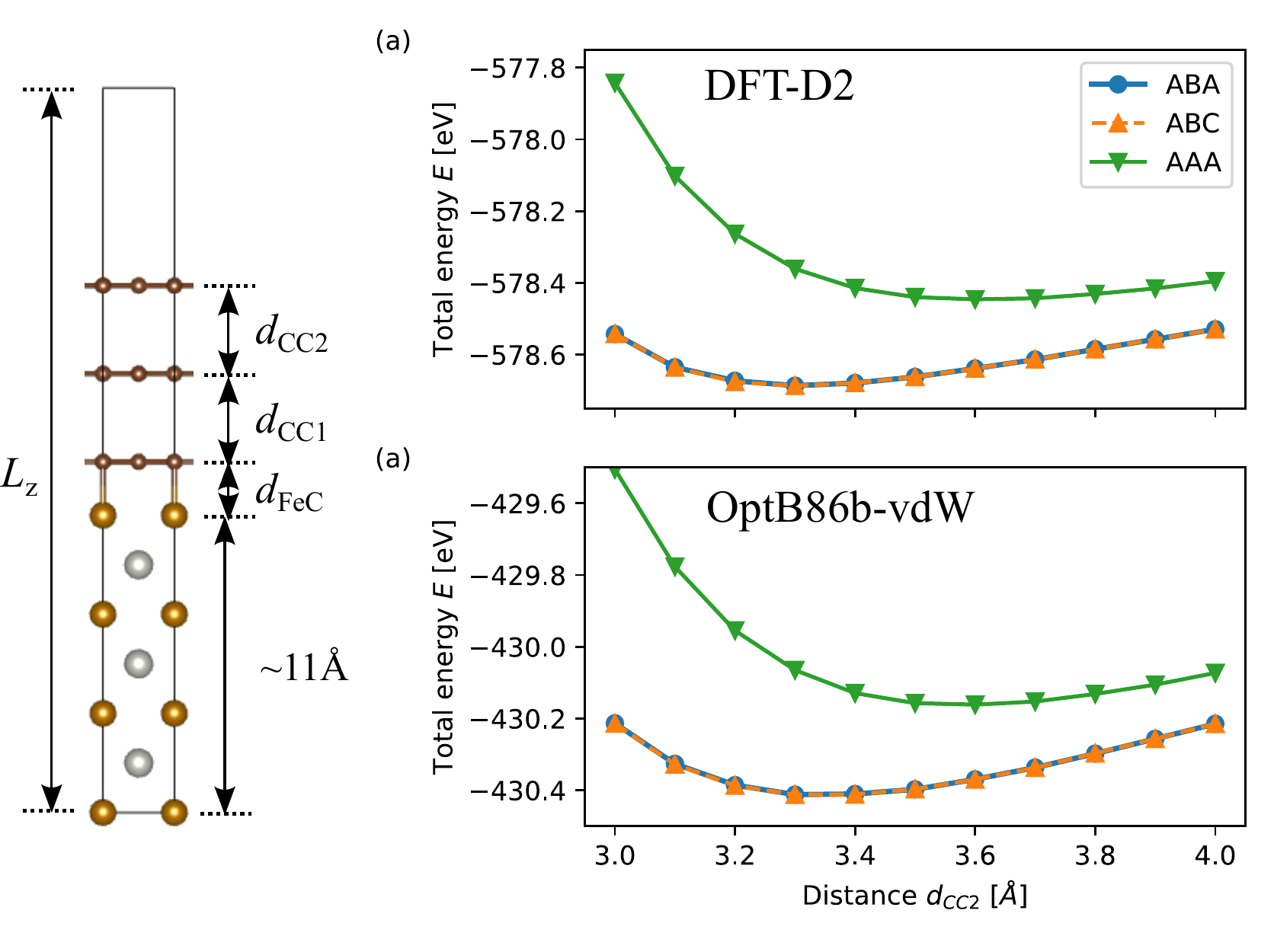}
    \caption{
         Total energy as a function of the graphene-graphene distance ($d_\mathrm{CC2}$) of ABA-, ABC-, and AAA-stacked trilayer graphenes on FePd(001) surface.
         Functionals of DFT-D2 (a) and OptB86b-vdW (b) are also tested.
    }
    \label{fig:fepdgr3dist}
\end{figure}

\begin{table}[h]
\caption{
Optimized interlayer distances of ABA- and ABC-stacked trilayer graphene on FePd(001).
The functionals of DFT-D2 and OptB86b-vdW (the values in parentheses) are tested.
}
\label{tbl:trilayer}
\begin{tabular}{ccrrr}
\toprule
Stacking order & Supercell size & \multicolumn{3}{c}{Interlayer distance} \\
 &  $L_z$~(\AA) & $d_\mathrm{FeC}$~(\AA) & $d_\mathrm{CC1}$~(\AA) & $d_\mathrm{CC2}$~(\AA)  \\
\midrule
 ABA & 26.1 & 1.97~(1.98) & 3.18~(3.23) & 3.24~(3.33) \\
     & 27.1 & 1.97~(1.98) & 3.18~(3.23) & 3.24~(3.33) \\
\midrule
 ABC & 26.1 & 1.97~(1.98) & 3.20~(3.21) & 3.30~(3.33) \\
     & 27.1 & 1.97~(1.98) & 3.20~(3.20) & 3.30~(3.33) \\
\bottomrule
\end{tabular}
\end{table}

Similar to the previous section,  we also investigate the structure of trilayer graphenes on FePd(001).
Figure~\ref{fig:fepdgr3dist} shows the supercell model and the total energy's depencence on the upper interlayer distance $d_\mathrm{CC2}$ representing the gap 
between the first and second highest graphene layers.
($L_z \approx 27~\text{\AA}$, $d_\mathrm{FeC} \approx 2~\text{\AA}$ and $d_\mathrm{CC1} \approx 3.3~\text{\AA}$ are assumed in this calculation.)
For both DFT-D2 and OptB86b-vdW, the behavior ABA and ABC stacked model are close and stable at $d_\mathrm{CC2}=3.2 \sim 3.4~\text{\AA}$.

Table~\ref{tbl:trilayer} shows interlayer distances of ABA- and ABC-stacked model after atomic position optimization.
The lower graphene-graphene distance $d_\mathrm{CC1}$ is about $3.2~\text{\AA}$, which is slightly smaller than that of the bulk graphite.
The upper distance $d_\mathrm{CC2}$ has a difference depending on the choice of the functionals used in the calculation.
Besides, the graphene-graphene distance of the ABC stacked model increases from that of the ABA.
The general trend of $d_\mathrm{CC1} < d_\mathrm{CC2}$ is commonly seen.
In addition, the total energy difference between ABA- and ABC- stacked structure is much smaller than  $0.001$~eV per single C atom, which is a negligibly small value.